\documentclass[twocolumn,floatfix,aps]{revtex4}
\usepackage{bm}
\usepackage{graphicx,color}

\usepackage{amsmath,amssymb}

\begin{document}

\title{Crossover between parabolic and hyperbolic scaling, oscillatory modes and resonances near flocking} 
\author{L L Bonilla$^{1,2}*$ and C. Trenado$^1$}
\affiliation{$^1$G. Mill\'an Institute for Fluid Dynamics, Nanoscience and
Industrial Mathematics, and Department of Materials Science and Engineering and Chemical Engineering, Universidad Carlos III de Madrid, 28911 Legan\'es,
Spain\\
$^2$Courant Institute for Mathematical Sciences, New York University, 251 Mercer St., New York, NY 10012\\
$^*$Corresponding author. E-mail: bonilla@ing.uc3m.es} 
\date{\today}
\begin{abstract}
A stability and bifurcation analysis of a kinetic equation indicates that the flocking bifurcation of the two-dimensional Vicsek model exhibits an interplay between parabolic and hyperbolic behavior. For box sizes smaller than a certain large value, flocking appears continuously from a uniform disordered state at a critical value of the noise. Because of mass conservation, the amplitude equations describing the flocking state consist of a scalar equation for the density disturbance from the homogeneous particle density (particle number divided by box area) and a vector equation for a current density. These two equations contain two time scales. At the shorter scale, they are a hyperbolic system in which time and space scale in the same way. At the longer, diffusive, time scale, the equations are parabolic. The bifurcating solution depends on the angle and is uniform in space as in the normal form of the usual pitchfork bifurcation. We show that linearization about the latter solution is described by a Klein-Gordon equation in the hyperbolic time scale. Then there are persistent oscillations with many incommensurate frequencies about the bifurcating solution, they produce a shift in the critical noise and resonate with a periodic forcing of the alignment rule. These predictions are confirmed by direct numerical simulations of the Vicsek model. 
\end{abstract}

\maketitle

\section{Introduction} \label{sec:1}
Flocking phenomena occurs in bacteria \cite{igo01,igo04,bal15}, spermatozoa \cite{cre16}, insects \cite{cav17,cav18}, birds \cite{ton95,ton05,bal08,att14}, animals \cite{cou05}, active gels \cite{sim02,jul07}, or interacting robots \cite{rub14}. Theoretical understanding of flocking benefits from analysis of simple models such as that proposed by Vicsek {\em et al} \cite{vic95}. In the Vicsek model (VM), the velocities of $N$ particles moving with equal speed in a box with periodic boundary conditions are updated so that the velocity of each particle adopts the direction of the average velocity of its close neighbors with some alignment noise ({\em conformist or majority rule}) \cite{vic95,vic12}. This system exhibits a phase transition from disordered to coherent behavior of the particles: when the alignment noise is sufficiently small or the particle density high enough, particles move coherently as a swarm.  Finite size effects are very important. Below a critical size of the box, flocking in the VM with forward update occurs as a continuous bifurcation from a disordered state with uniform particle density to an ordered state characterized by nonzero average speed of the particles \cite{vic95}. For box size larger than critical, the bifurcation is discontinuous and a variety of patterns are possible \cite{gre04,sol15}. Independently of the box size, the bifurcation is always continuous for VM with backward update \cite{bag09}. 

As the VM is straightforward to simulate numerically, many variations thereof have contributed to our understanding of flocking \cite{vic12,ram10,mar13,cav18}. To delve deeper into flock formation, many authors have derived continuum equations from the VM and its variants, often creating new models in the process (cf. the review papers Refs.~\cite{vic12,ram10,mar13,cav18}). Several authors have proposed kinetic theory equations based on the VM and then derived continuum equations from them. In a remarkable formulation, T. Ihle has derived several discrete-time kinetic equations that keep many features of the VM \cite{ihl11,ihl15,ihl16}. He then derived coupled continuum equations for the particle density and the momentum (or particle current) density by means of a Chapman-Enskog procedure valid near the transition to flocking \cite{ihl16}. These continuum equations contain terms that appear in the Toner-Tu macroscopic theory \cite{ton95}, and their coefficients have explicit expressions. However, Ihle's derivation introduces scaling {\em a posteriori} and it is not a systematic derivation based in bifurcation theory. 

In this paper, we analyze flocking in the two-dimensional (2D) VM by using systematically bifurcation theory for its Enskog kinetic equation \cite{ihl11}. A linear stability analysis of the time-independent uniform distribution shows that it becomes unstable at zero wavenumber when a real eigenvalue increases past 1. The corresponding eigenfunction is the complex first harmonic in the orientation of the particle velocity. The conservation of the number of particles implies that there is always a mode with eigenvalue 1 corresponding to particle density (more precisely, the difference between particle density and the homogeneous density, number of particles divided by box area, which we call {\em density disturbance}). Thus the amplitude equations produced by bifurcation theory correspond to coupled equations for the complex amplitude of the first harmonic and for the density disturbance. These complex-valued amplitude equations are equivalent to a system of real equations for the density disturbance and a current density. To leading order, these equations are hyperbolic: time and space scale with the same exponent (ballistic regime), and they become parabolic (diffusive) when first order corrections are added (time scales as the square of space). The scaling in the ballistic regime is akin to that observed in experiments of insect swarms \cite{att14}. 

Remarkably, the dissipation in the longer time scale drives the equations to a spatially uniform state that yields the standard diagram of a supercritical pitchfork bifurcation. However, near the bifurcation point, there are small-amplitude oscillations with precise frequencies given by solving a Klein-Gordon equation. Farther from the bifurcation point, the distinction between hyperbolic and parabolic scalings becomes blurred, dissipation dominates and these oscillatory modes disappear. One effect of the oscillatory modes is to shift the bifurcation point (critical noise) to smaller noise values by an amount proportional to the average of the density disturbance squared. A shift in the critical noise to smaller values is seen in direct numerical simulations of the VM. To confirm the existence of the oscillatory modes, we have added a harmonic forcing to the Vicsek alignment rule and directly simulated the resulting Vicsek model. As the frequency of the forcing resonates with the fundamental mode, the amplitude of the latter in the discrete Fourier transform of the polarization increases. Then the corresponding peak in the discrete Fourier transform becomes the largest one after the main peak of zero frequency whose height equals the static ensemble average of the polarization. If the forcing frequency is close to that of a higher mode, nearby modes with equal or lower frequencies are excited due to the nonlinearity of the amplitude equations. Thus, direct simulations of the VM confirm these predictions based on the bifurcation theory for the kinetic equation.

The rest of the paper is as follows. The VM and its nondimensionalization are presented in Section \ref{sec:2}. In Section \ref{sec:3}, we review the derivation of the kinetic equation for the VM in the limits of small and large particle density. The corresponding collision terms are binary and Enskog-like, respectively. The kinetic equation is discrete in time, nonlocal in space and strongly nonlinear. The number of particles is a conserved quantity for this equation. In Section \ref{sec:4}, we pose and study the linear stability of the disordered solution having time-independent uniform particle density. Disorder is unstable if at least one eigenvalue has modulus larger than one. As a consequence of conservation of the number of particles, one is always an eigenvalue corresponding to a constant eigenfunction. We have not solved the eigenvalue problem in the general case of nonzero wave number. However, perturbation theory for wave vectors of small modulus suggests that the eigenvalues decrease as the modulus of the wave vector increases, thereby supporting our choice of eigenvalues with zero wave number. If an eigenvalue corresponding to nonzero wave number exits first the unit circle, then the corresponding mode depends on space. This could be the case for the band patterns found in the literature. We have constructed the bifurcating solutions issuing from the disordered one by means of a Chapman-Enskog method described in Section \ref{sec:5} (cf. Refs.~\cite{bon00,ace05,BT10}). The analysis of the amplitude equations is presented in Sections \ref{sec:6} (spatially uniform solutions) and \ref{sec:7} (spatio-temporal solutions). A discussion of our results is presented in Section \ref{sec:8} and the Appendices are devoted to different technical matters.

\section{2D Vicsek Model} \label{sec:2}
We consider an angular noise Vicsek model with forward updating rule. Our choice differs from Vicsek's \cite{vic95} in the updating rule and it is the same as in Ref.~\cite{ihl16}. See Ref.~\cite{bag09} for a discussion on how different definitions of the VM affect the character of the order-disorder phase transition. 

More specifically, in dimensional units, $N$ particles with positions $\mathbf{x}_j$ and velocities $\mathbf{v}_j=v_0(\cos\theta_j,\sin\theta_j)$, $j=1,\ldots,N$, are inside a square box of size $L$ and we use periodic boundary conditions. Initially, the particle positions are random. The particles undergo discrete dynamics so that their positions are forwardly updated, 
\begin{eqnarray}
\mathbf{x}_j(t+\tau)=\mathbf{x}_j(t)+ \tau\mathbf{v}_j(t+\tau).\label{eq1}
\end{eqnarray}
Here $t=0,\tau, 2\tau,\ldots$. The angle of a particle $i$ is updated according to the Vicsek angular noise rule 
\begin{eqnarray}
\theta_i(t+\tau)=\mbox{Arg}\left(\sum_{|\mathbf{x}_j-\mathbf{x}_i|<R_0}e^{i\theta_j(t)}\right)+\xi_i(t),\label{eq2}
\end{eqnarray}
where we sum over all particles that, at time $t$, are inside a circle of radius $R_0$ centered at $\mathbf{x}_i$ (the {\em circle of influence} or {\em interaction circle}). The sum includes the particle $i$. At each time, $\xi_i(t)$ is a random number chosen with probability density $g(\xi)$. Typically, $g(\xi)$ is uniform inside an interval $(-\eta/2,\eta/2)$:
\begin{equation}
g(\xi)=\left\{\begin{array}{cc}
\frac{1}{\eta}, & |\xi|<\frac{\eta}{2},\\
0, &\mbox{otherwise,}\\
\end{array}\right. \label{eq3}
\end{equation}
where $0\leq\eta\leq 2\pi$. 

We nondimensionalize the model according to Table \ref{table1}. 

\begin{table}[ht]
\begin{center}\begin{tabular}{|c|c|c|c|c|}
 \hline
$\mathbf{x}$, $R_0$, $L$& $\mathbf{v}$ & $t$ &$\theta$, $\xi$\\ \hline
$v_0\tau$ & $v_0$ & $\tau$ & --\\ 
\hline
\end{tabular}
\end{center}
\caption{Units for nondimensionalizing the equations of the model. }
\label{table1}
\end{table}

In our nondimensional units, the VM is described by Eq.~\eqref{eq2} (with nondimensional time, space and $R_0$) and 
\begin{equation}
\mathbf{x}_j(t+1)=\mathbf{x}_j(t)+(\cos\theta_j(t+1),\sin\theta_j(t+1)), \label{eq4}
\end{equation}
which is Eq.~\eqref{eq1} with $v_0=1$, $\tau=1$. In these units, the nondimensional average particle density becomes
\begin{eqnarray}
\rho_0=\frac{Nv_0^2\tau^2}{L^2}, \label{eq5}
\end{eqnarray}
whereas the average number of neighbors of a particle, $M=N\pi R_0^2/L^2$, remains an unchanged dimensionless parameter.

 Collective consensus is quantified by the complex order parameter 
\begin{eqnarray}
Z=W\, e^{i\Upsilon}=\frac{1}{N}\sum_{j=1}^N e^{i\theta_j}, \label{eq6}
\end{eqnarray}
whose amplitude $0<W<1$ (polarization) measures macroscopic coherence of the particles and $\Upsilon$ is their average phase. 

\section{Kinetic theory} \label{sec:3}
Here, we follow Ihle's work and derive a kinetic equation in the limit of many particles ($N\gg 1$) by assuming molecular chaos \cite{ihl11,ihl16}. Let $f(\theta,\textbf{x},t)\, d\textbf{x}\, d\theta$ be the number of particles in an area $d\textbf{x}$ centered at position $\mathbf{x}$ that move into a direction between $\theta$ and $\theta+d\theta$ at time $t$. Assuming that all particles are independent and identically distributed before undergoing a collision described by Eqs.~\eqref{eq2}-\eqref{eq4}, the $N$-particle probability density is
\begin{equation}
P_N(\mathbf{x}_1,\theta_1,\ldots,\mathbf{x}_N,\theta_N,t)=\prod_{i=1}^N\frac{f(\mathbf{x}_j,\theta_j,t)}{N}. \label{eq7}
\end{equation}
Here $f(\mathbf{x}_j,\theta_j,t)$ and $f(\mathbf{x}_j,\theta_j,t)/N$, $j=1,\ldots, N$ are the one-particle distribution functions and probability densities, respectively. Eq.~\eqref{eq7} is the assumption of molecular chaos first introduced by Boltzmann when deriving his transport equation \cite{hua87}. The distribution function after the collision is found by integrating the $N$-particle probability density at time $t+1$ over $\mathbf{x}_2,\ldots,\mathbf{x}_N$ and dividing by $N$. The result is \cite{ihl16}
\begin{eqnarray}
f(\textbf{x}+\textbf{v},\theta,t+1)=C[f](\theta,\textbf{x},t),\quad\textbf{v}=(\cos\theta,\sin\theta), \label{eq8}
\end{eqnarray}
\begin{widetext}
\begin{eqnarray}
&&C[f]\!=\!\int^{\pi}_{-\pi}\! d\xi \, g(\xi)\sum_{n=1}^N\left(\begin{array}{c} N-1\\ n-1 \\ \end{array}\right)
\int_{[-\pi,\pi]^n} \hat{\delta}(\theta-\xi-\Phi_1(\tilde{\theta}_1,\ldots,\tilde{\theta}_n))\, f(\textbf{x},\tilde{\theta}_1,t)\left(1-\frac{M_R(\mathbf{x},t)}{N}\right)^{N-n}\nonumber\\
&&\times \prod_{i=2}^{n}\!\left[\int_{|\mathbf{x}_i-\mathbf{x}|<R_0}\frac{f(\textbf{x}_i,\tilde{\theta}_i,t)}{N} d\tilde{\theta}_i d\textbf{x}_i\right]\, d\tilde{\theta}_1,\label{eq9}\\
 &&M_R(\textbf{x},t)= \int_{|\mathbf{x}'-\mathbf{x}|<R_0}\rho(\textbf{x}',t)\, d\mathbf{x}',\quad \rho(\mathbf{x},t)=\int_{-\pi}^\pi f(\mathbf{x},\theta,t)\, d\theta,\quad\int\rho(\mathbf{x},t)d\mathbf{x}=N,\label{eq10}\\
&& \Phi_1(\tilde{\theta}_1,\ldots,\tilde{\theta}_n)=\mbox{Arg}\left(\sum_{|\mathbf{x}_j-\mathbf{x}_1|<R_0}e^{i\tilde{\theta}_j}\right)\!=\mbox{Arg}\left(\sum_{j=1}^n e^{i\tilde{\theta}_j}\right)\!.\label{eq11} 
\end{eqnarray}
\end{widetext}
Here $\Phi_1$ given by Eq.~\eqref{eq11}, is the average direction of the vector sum of all particle velocities (including particle 1) inside the circle of radius $R_0$. $\hat{\delta}(x)=\sum_{l=-\infty}^\infty\delta(x+2\pi l)$ is a periodized delta function that incorporates the collision rule and the integral over $\xi$ averages over the noises. In Eq.~\eqref{eq9}, $n$ is the number of particles inside the interaction circle about particle 1 (the latter included). The average number of particles inside a circle of radius $R_0$ about position $\mathbf{x}$ is $M_R(\textbf{x},t)$, given by Eq.~\eqref{eq10}. The combinatorial factor in Eq.~\eqref{eq9} counts the number of possible selections of neighbors of particle 1 (excluding the latter) out of the $N-1$ other particles. The factor $(1-M_R/N)^{N-n}$ in Eq.~\eqref{eq9} gives the probability that the particles $n+1,\ldots,N$ are not inside the interaction circle of particle 1. The factor $\prod_{i=1}^n\int_{|\mathbf{x}_i-\mathbf{x}|<R_0}f(\textbf{x}_i,\theta_i,t) d\mathbf{x}_i$ is the probability that particles $2,\ldots, n$ be within interaction distance of particle 1 times their angular distribution, given that they are within its interaction circle. When we integrate Eq.~\eqref{eq9} over $\theta$, we find that the particle density immediately after collisions equals that before:
\begin{equation}
\int C[f](\theta,\textbf{x},t)\, d\theta=\rho(\mathbf{x},t). \label{eq12}
\end{equation}

We may adopt two opposite approximations of the collision operator \eqref{eq9}. For very diluted particle ensembles having small average density, $\rho_0=N/L^2$, terms with $n\geq 2$ in Eq.~\eqref{eq9} provide negligible contributions. Then we get a binary collision operator
\begin{eqnarray}
C_B[f]\!=\!\int^{\pi}_{-\pi}\! d\xi \frac{g(\xi)}{1+M_R}\!\left[\int^{\pi}_{-\pi}\! \hat{\delta}(\theta-\xi-\tilde{\theta}_1)f(\textbf{x},\tilde{\theta}_1,t) d\tilde{\theta}_1\right.\nonumber\\
+ \int^{\pi}_{-\pi}\!\int^{\pi}_{-\pi}\! \hat{\delta}(\theta-\xi-\Phi_1(\tilde{\theta}_1,\tilde{\theta}_2))\, f(\textbf{x},\tilde{\theta}_1,t)\quad\nonumber\\
\times\left.\left(\int_{|\mathbf{x}_2-\mathbf{x}|<R_0}f(\textbf{x}_2,\tilde{\theta}_2,t) d\textbf{x}_2\right)\! d\tilde{\theta}_2 d\tilde{\theta}_1\right]\!,\quad \label{eq13}
\end{eqnarray}
with $M_R=M_R(\mathbf{x},t)$, which has been normalized so that $\int^{\pi}_{-\pi}\! C_B[f]d\theta=\rho(\mathbf{x},t)$.

Secondly, for larger densities and $n/N\ll 1$ as $N\to\infty$, the combinatorial factor times $(1-M_R/N)^{N-n}$ becomes
\begin{equation*}
\frac{(N-1)!}{(n-1)!(N-n)!}\left(1-\frac{M_R}{N}\right)^{N-n}\sim \frac{N^{n-1}}{(n-1)!}\, e^{-M_R},
\end{equation*}
and \eqref{eq9} produces an Enskog-type collision operator \cite{ihl16}
\begin{widetext}
\begin{eqnarray}
C_E[f]\!=\!\int^{\pi}_{-\pi}\! d\xi \, g(\xi) e^{-M_R(\mathbf{x},t)}\!\sum_{n=1}^{\infty}\int_{[-\pi,\pi]^n}\!\!\frac{\hat{\delta}(\theta-\xi-\Phi_1(\tilde{\theta}_1,\ldots,\tilde{\theta}_n))}{(n-1)!}
 f(\textbf{x},\tilde{\theta}_1,t)\prod_{i=2}^{n}\!\left[\int_{|\mathbf{x}_i-\mathbf{x}|<R_0}f(\textbf{x}_i,\tilde{\theta}_i,t) d\tilde{\theta}_i d\textbf{x}_i\right]\! d\tilde{\theta}_1. \label{eq14}
 \end{eqnarray}
 \end{widetext}
For active particles in a disordered state, the density $\rho(\mathbf{x},t)$ equals the constant average density, $\rho_0=N/L^2$, and the uniform distribution function, $f_0=\rho_0/(2\pi)$, is a fixed point of the collision operators: 
\begin{eqnarray}
C[f_0]=f_0,\quad C_B[f_0]=f_0,\quad C_E[f_0]=f_0. \label{eq15}
\end{eqnarray}
Henceforth, we shall use the Enskog collision operator \eqref{eq14}.

\section{Linear stability} \label{sec:4}
To analyze the order-disorder transition we linearize the kinetic equation using $f=f_0+\epsilon \tilde{f}(\theta,\textbf{x},t)$, $\epsilon\ll 1$, thereby obtaining
\begin{widetext}
\begin{eqnarray}
\tilde{f}(\theta,\textbf{x}+\mathbf{v},t+1)=\sum_{n=1}^\infty\frac{e^{-M}}{(n-1)!}\left(\frac{M}{2\pi}\right)^{n-1}\!\! \int^{\pi}_{-\pi}\! d\xi\, g(\xi)\!\int_{[-\pi,\pi]^n}\hat{\delta}(\theta-\xi-\Phi_1)\!\left[\tilde{f}(\mathbf{x},\tilde{\theta}_1,t)+\frac{n-1}{\pi R_0^2}\! \int_{|\mathbf{x}'-\mathbf{x}|<R_0}\!\tilde{f}(\mathbf{x}',\tilde{\theta}_1,t)d\mathbf{x}' \right]\!  \nonumber\\
\times\prod_{l=1}^n d\tilde{\theta}_l -\frac{M}{2\pi}\frac{1}{\pi R_0^2}\int_{|\mathbf{x}'-\mathbf{x}|<R_0}\int^{\pi}_{-\pi}\tilde{f}(\theta',\textbf{x}',t) d\textbf{x}' d\theta'.\quad \label{eq16}
\end{eqnarray}
\end{widetext}
Here $M=\rho_0\pi R_0^2$ and $\Phi_1(\tilde{\theta}_1,\ldots,\tilde{\theta}_n)$ has been defined in Eq.~\eqref{eq11}. The separation of variables ansatz $\tilde{f}(\mathbf{x},\theta,t)=\tilde{F}(\mathbf{x},\theta)\, h(t)$ produces a discrete equation $h(t+1)/h(t)=Q$, where $Q$ is the separation constant. Thus $h(t)=Q^t$. The equation for $\tilde{F}$ is an eigenvalue problem that yields $Q$. Moreover, $\tilde{F}$ is a periodic function of space and it can be written as a Fourier series expansion in plane waves, $e^{i\mathbf{K}\cdot\mathbf{x}}$, in which the components of the wave vectors are integer multiples of $2\pi/L$. In the limit as $L\to\infty$, the wave vectors $\mathbf{K}$ are real valued and the Fourier series becomes a Fourier integral. Setting $\tilde{F}=e^{i\mathbf{K}\cdot\mathbf{x}}\varphi(\theta;\mathbf{K})$, we are led to the separation of variables ansatz $\tilde{f}=Q^t e^{i\mathbf{K}\cdot\mathbf{x}}\varphi(\theta)$, where $Q$ and $\varphi(\theta)$ are both functions of $\mathbf{K}$. This procedure of separation of variables is typically used in discussions of the Fourier-von Neumann stability of finite difference numerical methods for linear partial differential equations; see Ref.~\cite{hab13}. From Eq.~\eqref{eq16}, the integration of the plane wave on the disk of radius $R_0$ yields the eigenvalue problem for $\varphi(\theta)$:
\begin{eqnarray}
Qe^{i\mathbf{K}\cdot\mathbf{v}}\varphi- C^{(1)}[\varphi]=0,\label{eq17}
\end{eqnarray}
\begin{widetext}
\begin{eqnarray}
\!\!\!C^{(1)}[\varphi]\!=\!\frac{2J_1(|\mathbf{K}|R_0)}{|\mathbf{K}|R_0}\left[\sum_{n=1}^\infty\frac{e^{-M}}{(n-1)!}\left(\frac{M}{2\pi}\right)^{n-1}\!\! \left(n-1+\frac{|\mathbf{K}|R_0}{2J_1(|\mathbf{K}|R_0)}\right)\!\!\int^{\pi}_{-\pi}\! d\xi\, g(\xi)\!
\!\int_{[-\pi,\pi]^n} \hat{\delta}(\theta-\xi-\Phi_1)\varphi(\tilde{\theta}_1)\!\prod_{l=1}^n d\tilde{\theta}_l\!\right.\nonumber\\
\left. -\frac{M}{2\pi}\int^{\pi}_{-\pi}\varphi(\tilde{\theta}) d\tilde{\theta}\right]\!.  \label{eq18}
\end{eqnarray}
We have $C^{(1)}[1]=1$, and therefore the uniform distribution $f_0=\rho_0/(2\pi)$ solves Eq.~\eqref{eq17} with  $|\mathbf{K}|=0$ and $Q=1$. 

We now seek non-constant solutions of Eq.~\eqref{eq17} by inserting the Fourier expansion $\varphi(\theta)=\sum_{j=-\infty}^\infty\varphi_j e^{ij\theta}$. We find
\begin{eqnarray}
\sum_{j=-\infty}^\infty\left[Q(e^{i\mathbf{K}\cdot(\cos\theta,\sin\theta)})_j-C^{(1)}[\varphi]_j\right] e^{ij\theta}=0, \label{eq19}
\end{eqnarray}
from which we obtain the eigenvalue problem:
\begin{eqnarray}
\sum_{l=-\infty}^\infty \{C^{(1)}[e^{ij\theta}]_j\delta_{jl}-Q(e^{i\mathbf{K}\cdot(\cos\theta,\sin\theta)+ij\theta})_l\} \varphi_{l}=0. \label{eq20}
\end{eqnarray}
Here the subscripts $j$ and $l$ indicate that $f(\theta)_j$ and $f(\theta)_l$ are the coefficients of the respective harmonics in the Fourier series of the function $f(\theta)$, and we have used $C^{(1)}[e^{il\theta}]_j=0$ for $j\neq l$ \cite{ihl16}. Equivalently, $1/Q$ are the eigenvalues of a matrix $\mathcal{M}_{jl}$: 
\begin{eqnarray}
&&\mathcal{M}_{jl}=\frac{(e^{i\mathbf{K}\cdot(\cos\theta,\sin\theta)+ij\theta})_l}{C^{(1)}[e^{ij\theta}]_j}, \quad (e^{i\mathbf{K}\cdot(\cos\theta,\sin\theta)+ij\theta})_j =(e^{i\mathbf{K}\cdot(\cos\theta,\sin\theta)} )_0=J_0(|\mathbf{K}|), \nonumber\\
&&C^{(1)}[e^{ij\theta}]_j=\!\left(\int^{\pi}_{-\pi}\! e^{-ij\xi} g(\xi)\,d\xi\right)\!\sum_{n=1}^\infty\frac{M^{n-1}e^{-M}}{(n-1)!}\!\!\left[(n-1)\frac{2J_1(|\mathbf{K}|R_0)}{|\mathbf{K}|R_0}+1\right]\!\int_{[-\pi,\pi]^n}e^{ij(\tilde{\theta}_1-\Phi_1)}\prod_{l=1}^n \frac{d\tilde{\theta}_l}{2\pi} .  \label{eq21}
\end{eqnarray}
If $\mathbf{K}=K(0,1)$, the off-diagonal matrix elements are $J_{l-j}(K)/C^{(1)}[e^{ij\theta}]_j$.
\end{widetext}

\subsection{Space independent eigenfunctions}
In this paper, we study solutions that bifurcate from disorder with zero wave number, which correspond to bifurcations for box sizes below critical, see Section \ref{sec:2}. For $|\mathbf{K}|=0$, Eq.~\eqref{eq20} produces the following eigenvalues and eigenfunctions:
\begin{eqnarray}
Q_j=C^{(1)}[e^{ij\theta}]_j, \quad\varphi_j(\theta)=e^{ij\theta},\quad (\varphi_j)_l=\delta_{lj},\label{eq22}
\end{eqnarray}
with $j,l=1,2,\ldots$. The disordered state is stable when $|Q_j|\leq 1$ for all $j$, and unstable if $|Q_j|>1$ for some $j$. $\varphi_0=1$ is one eigenfunction corresponding to eigenvalue $Q_0=1$. The eigenvalue with largest modulus for $j\neq 0$ is $Q_1$, which, for large $M$, becomes \cite{ihl16}
\begin{eqnarray}
Q_1\!&\sim& \frac{\sqrt{\pi M}}{2}\int^{\pi}_{-\pi}\! e^{-i\xi} g(\xi)d\xi=\frac{\sqrt{\pi M}}{\eta}\sin\frac{\eta}{2}.  \label{eq23}
\end{eqnarray}
Other eigenvalues have moduli smaller than 1 in the limit as $M\to\infty$, as shown in Appendix \ref{ap:2}. 

\begin{figure}[h]
\begin{center}
\includegraphics[width=8cm]{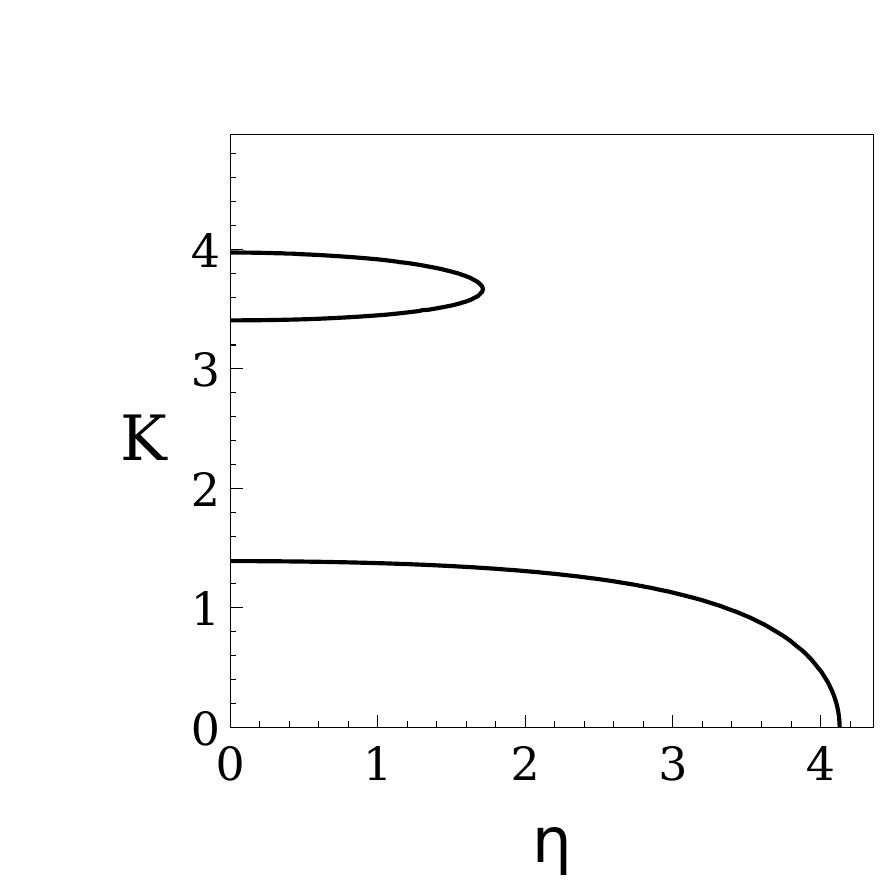}
\end{center}
\caption{Critical wave vector $K=|\mathbf{K}|$ versus $\eta$ for $M=7$ and $\rho_0=5$ obtained by solving $Q_1=1$, where $Q_1$ is given by Eqs.~\eqref{eq21} and \eqref{eq26}. \label{fig1}}
\end{figure}

\subsection{Perturbation of eigenvalues for small off-diagonal elements}
We do not know how to find the eigenvalues of the matrix $\mathcal{M}_{jl}$, given by Eq.~\eqref{eq21}, for general nonzero $\mathbf{K}$. However, the diagonal entries $\mathcal{M}_{jj}$ are proportional to $J_0(|\mathbf{K}|)=1+O(|\mathbf{K}|)$ (as $|\mathbf{K}|\to 0$), whereas the off-diagonal elements of $\mathcal{M}_{jl}$ with $j\neq l$ vanish for zero wave vector. Thus, for small $|\mathbf{K}|$, the off-diagonal elements of the matrix $\mathcal{M}_{l,j}$ of Eq.~\eqref{eq21} are small compared to the diagonal elements. Assuming that the matrix $\mathcal{M}_{jl}$ is equal to the matrix of its diagonal elements plus a small perturbation, we can use regular perturbation theory to calculate its eigenvalues. The situation is analogous to the usual perturbation theory of eigenvalues in non-relativistic Quantum Mechanics as explained in Ref~\cite{LL3}. The first order correction to the eigenvalues of $\mathcal{M}_{jl}$ is given by the diagonal elements of the perturbation matrix. However, this perturbation matrix is $\mathcal{M}_{jl}$ minus its diagonal, and therefore it has zero diagonal elements. Thus, we need to calculate the eigenvalues by using second-order perturbation theory for a perturbation matrix comprising the off-diagonal elements of $\mathcal{M}_{jl}$. We obtain [cf. Eq.~(38.10) of Ref~\cite{LL3}]: 
\begin{eqnarray}
\frac{1}{Q_j}=\frac{J_0(|\mathbf{K}|)}{C^{(1)}[e^{ij\theta}]_j}- \sum_{l\neq j}\frac{\mathcal{M}_{l,j}\mathcal{M}_{j,l}}{\frac{J_0(|\mathbf{K}|)}{C^{(1)}[e^{ij\theta}]_l}-\frac{J_0(|\mathbf{K}|)}{C^{(1)}[e^{ij\theta}]_j}}, \label{eq24}
\end{eqnarray}
which holds for $0\leq |\mathbf{K}|<\gamma_{1,0}$ [$\gamma_{1,0}\approx 2.4048$ is the first zero of the Bessel function $J_0(x)$]. We are interested in the eigenvalue close to $1/Q_1$, because $Q_1$ has the largest modulus for $|\mathbf{K}|=0$. It is approximately given by
\begin{eqnarray}
\frac{1}{Q_1}\approx\frac{J_0(|\mathbf{K}|)}{C^{(1)}[e^{i\theta}]_1}- \frac{\mathcal{M}_{2,1}\mathcal{M}_{1,2}}{\frac{J_0(|\mathbf{K}|)}{C^{(1)}[e^{i2\theta}]_2}-\frac{J_0(|\mathbf{K}|)}{C^{(1)}[e^{i\theta}]_1}}, \label{eq25}
\end{eqnarray}
in which we have ignored higher order terms having $|l-1|>1$. For $\mathbf{K}=K(0,1)$, these terms are proportional to $[J_{l-1}(K)]^{2}=O(|\mathbf{K}|^{2(l-1)})$ (with $l>2$). Thus, compared with the last term in Eq.~\eqref{eq25}, which is $O(|\mathbf{K}|^{2})$, they can be ignored in the limit as $|\mathbf{K}|\to 0$. This also occurs for general $\mathbf{K}$. We have $(e^{i\mathbf{K}\cdot(\cos\theta,\sin\theta)})_0=J_0(|\mathbf{K}|)$, and $(e^{i\mathbf{K}\cdot(\cos\theta,\sin\theta)+i\theta})_0=((\cos\theta+i\sin\theta)(e^{i\mathbf{K}\cdot(\cos\theta,\sin\theta)})_0$ produces
\begin{eqnarray*}
&&\mathcal{M}_{2,1}=-\frac{1}{C^{(1)}[e^{i2\theta}]_2}(i\frac{\partial}{\partial K_x}-\frac{\partial}{\partial K_y})J_0(|\mathbf{K}|)\nonumber\\
&&\quad\quad=-\frac{i K_x- K_y}{|\mathbf{K}|\, C^{(1)}[e^{i2\theta}]_2}J_1(|\mathbf{K}|)\Longrightarrow\nonumber\\
&&\mathcal{M}_{1,2}\mathcal{M}_{2,1}=-\frac{[J_1(|\mathbf{K}|)]^2}{C^{(1)}[e^{i\theta}]_1C^{(1)}[e^{i2\theta}]_2}.
\end{eqnarray*}
Thus, Eq.~\eqref{eq25} becomes
\begin{eqnarray}
Q_1\approx\frac{C^{(1)}[e^{i\theta}]_1}{J_0(|\mathbf{K}|)+\frac{[J_1(|\mathbf{K}|)]^2C^{(1)}[e^{i\theta}]_1}{J_0(|\mathbf{K}|)\, (C^{(1)}[e^{i\theta}]_1-C^{(1)}[e^{i2\theta}]_2)}}. \label{eq26}
\end{eqnarray}
According to Eqs.~\eqref{eq21} and \eqref{eq26}, $Q_1=Q_1(\eta,|\mathbf{K}|)$. Then the equation $Q_1=1$ may have different solution branches $\eta(|\mathbf{K}|)$ for nonzero wave number, depending on the parameters $M$ and $\rho_0$. Two such branches are displayed in Fig.~\ref{fig1}. The lowest branch depicts the solution branch that prolongs the zero wave number solution given by Eq.~\eqref{eq23}. It satisfies $0\leq\eta(|\mathbf{K}|)<\eta(0)$ with $0<|\mathbf{K}|<2$, thereby justifying that the largest value of the multiplier $Q$ for this branch is attained at zero wave number. The possible upper branches can appear for $|\mathbf{K}|>\gamma_{1,0}\approx 2.4048$, roughly coinciding with intervals where $[J_1(|\mathbf{K}|)]^2<[J_0(|\mathbf{K}|)]^2$. The largest value of $\eta$ for the upper branch in Fig.~\ref{fig1} is smaller than $\eta(0)$. Thus, within the approximations we have made, the largest value of $Q$ occurs at zero wave number. 

\section{Bifurcation theory} \label{sec:5}
As justified in Section \ref{sec:4}, the largest multiplier is $Q_1=1$ corresponding to $\mathbf{K}=0$. The solution of the linearized equation 
\begin{eqnarray}
\mathcal{L}f^{(1)}\!\!\equiv\! f^{(1)}(\theta,t+1,\textbf{X},T)\!-\! C^{(1)}[ f^{(1)}](\theta,t,\textbf{X},T)\!=\!0,\label{eq27}
\end{eqnarray}
is  
\begin{eqnarray}
 f^{(1)}(\theta,t,\textbf{X},T,\epsilon)\!=\!\frac{r(\textbf{X},T;\epsilon)}{2\pi}\!+\!A(\textbf{X},T;\epsilon) e^{i\theta} \!+\!\mbox{cc},\label{eq28}\\
\textbf{X}=\epsilon\textbf{x},\quad T=\epsilon t.\quad \label{eq29}
\end{eqnarray}
Here cc means the complex conjugate of the preceding term. We do not need to include more terms in \eqref{eq28} because the other modes decay rapidly in the fast time scale $t$. The first term in Eq.~\eqref{eq28} is a space dependent disturbance of the uniform density, whereas the complex amplitude of the second term corresponds to a vector current density, as we will show below in Eqs.~\eqref{eq42}-\eqref{eq44}. 

We anticipate crossover scalings and therefore we shall use the Chapman-Enskog method \cite{bon00,ace05,BT10,neu15}. The Chapman-Enskog ansatz is \cite{BT10,bon00,ace05}, 
\begin{eqnarray}
f(\theta,\textbf{x},t;\epsilon)=f_0+ \epsilon f^{(1)}+\sum_{j=2}^\infty\epsilon^j f^{(j)}(\theta,t;r,A,\overline{A}), \label{eq30}\\
\frac{\partial r}{\partial T}=\mathcal{R}^{(0)}(r,A,\overline{A})+\epsilon\mathcal{R}^{(1)}(r,A,\overline{A})+O(\epsilon^2),\label{eq31}\\
\frac{\partial A}{\partial T}=\mathcal{A}^{(0)}(r,A,\overline{A})+\epsilon\mathcal{A}^{(1)}(r,A,\overline{A})+O(\epsilon^2),\label{eq32}
\end{eqnarray}
where $\overline{A}$ is the complex conjugate of $A$. We select a scaling  $\eta=\eta_c+\epsilon^2\eta_2$, which is appropriate for the case of the pitchfork bifurcation that occurs for space independent solutions. We will explain later what happens for a different choice of scaling. Inserting Eqs.~\eqref{eq30}-\eqref{eq32} into Eqs.~\eqref{eq8} and \eqref{eq14}, we obtain the following hierarchy of equations
\begin{eqnarray}
\mathcal{L}f^{(2)}=C^{(2)}[f^{(1)},f^{(1)}]-\textbf{v}\!\cdot\!\nabla_X f^{(1)}- \frac{\mathcal{R}^{(0)}}{2\pi}\nonumber\\
-\mathcal{A}^{(0)} e^{i\theta}+ \mbox{cc},\label{eq33}
\end{eqnarray}
\begin{eqnarray}
&&\mathcal{L}f^{(3)}=C^{(3)}[f^{(1)},f^{(1)},f^{(1)}]+2 C^{(2)}[f^{(1)},f^{(2)}]  \nonumber\\
&&\quad -\textbf{v}\!\cdot\!\nabla_X f^{(2)}-\frac{\mathcal{R}^{(1)}}{2\pi}-\mathcal{A}^{(1)} e^{i\theta}+ \mbox{cc}
\nonumber\\
&&\quad-\frac{1}{2}\!\left(\frac{\partial}{\partial T}+\textbf{v}\!\cdot\!\nabla_X\right)^2f^{(1)} +\eta_2\frac{\partial}{\partial\eta}C^{(1)}[f^{(1)}],\label{eq34}
\end{eqnarray}
etc. In these equations, $\mathcal{L}$ is given by Eqs.~\eqref{eq27} and \eqref{eq18}, and 
\begin{widetext}
\begin{eqnarray}
&&C^{(2)}[\varphi,\varphi]=\frac{\pi R_0^2}{2}\!\left[\sum_{n=2}^\infty\frac{ne^{-M}}{(n-2)!}\!\left(\frac{M}{2\pi}\right)^{n-2}\!\int_{-\pi}^\pi d\xi g(\xi)\int_{[-\pi,\pi]^n}\hat{\delta}(\theta-\xi-\Phi_1)\varphi(\tilde{\theta}_1)\varphi(\tilde{\theta}_2)\prod_{l=1}^n d\tilde{\theta}_l-2\left(\int_{-\pi}^\pi \varphi(\theta_1)d\theta_1\right)\!\right.\nonumber\\
&&\left.\!\times\!\sum_{n=1}^\infty\frac{ne^{-M}}{(n-1)!}\!\left(\frac{M}{2\pi}\right)^{n-1}\!\int_{-\pi}^\pi d\xi g(\xi)\int_{[-\pi,\pi]^n}\hat{\delta}(\theta-\xi-\Phi_1)\varphi(\tilde{\theta}_1)\prod_{l=1}^n d\tilde{\theta}_l
+\frac{M}{2\pi}\left(\int_{-\pi}^\pi \varphi(\theta_1)d\theta_1\right)^2\right]\!,\label{eq35}
\end{eqnarray}
\begin{eqnarray}
&&C^{(3)}[\varphi,\varphi,\varphi]=\frac{\pi^2R_0^4}{6}\!\left[\sum_{n=3}^\infty\frac{ne^{-M}}{(n-3)!}\!\left(\frac{M}{2\pi}\right)^{n-3}\!\int_{-\pi}^\pi d\xi g(\xi)\int_{[-\pi,\pi]^n}\hat{\delta}(\theta-\xi-\Phi_1)\varphi(\tilde{\theta}_1)\varphi(\tilde{\theta}_2)\varphi(\tilde{\theta}_3)\prod_{l=1}^n d\tilde{\theta}_l\right.\nonumber\\
&&-3\!\left(\int_{-\pi}^\pi \varphi(\theta_3)d\theta_3\right)\!\sum_{n=2}^\infty\frac{ne^{-M}}{(n-2)!}\!\left(\frac{M}{2\pi}\right)^{n-2}\!\int_{-\pi}^\pi d\xi g(\xi)\int_{[-\pi,\pi]^n}\hat{\delta}(\theta-\xi-\Phi_1)\varphi(\tilde{\theta}_1)\varphi(\tilde{\theta}_2)\prod_{l=1}^n d\tilde{\theta}_l\nonumber\\
&&+3\!\left(\int_{-\pi}^\pi \varphi(\theta_2)d\theta_2\right)^2\!\sum_{n=1}^\infty\frac{ne^{-M}}{(n-1)!}\!\left(\frac{M}{2\pi}\right)^{n-1}\!\!\int_{-\pi}^\pi d\xi g(\xi)\!\int_{[-\pi,\pi]^n}\!\!\hat{\delta}(\theta-\xi-\Phi_1)\varphi(\tilde{\theta}_1)\!\prod_{l=1}^n\! d\tilde{\theta}_l\!
-\!\left.\frac{M}{2\pi}\left(\int_{-\pi}^\pi \varphi(\theta_1)d\theta_1\right)^3\right]\!,\,\,\label{eq36}
\end{eqnarray}
and so on. Note that $C_E[f_0+\epsilon\tilde{\rho}]=f_0+\epsilon\tilde{\rho}$ and $C^{(1)}[\tilde{\rho}]=\tilde{\rho}$ for constant $\tilde{\rho}$ imply $C^{(2)}[1,1]=C^{(3)}[1,1,1]=0$, which can be checked from Eqs.~\eqref{eq35}-\eqref{eq36}. The solvability conditions for non-homogeneous equations of the hierarchy is that their right hand sides be orthogonal to the solutions of the homogeneous equation $\mathcal{L}\varphi=0$, namely 1 and $e^{i\theta}$, using the scalar product
\begin{equation}
\langle f(\theta),g(\theta)\rangle= \int_{-\pi}^\pi \overline{f(\theta)} g(\theta)\, d\theta. \label{eq37}
\end{equation}

We now proceed to derive the amplitude equations. We insert Eq.~\eqref{eq28} into Eq.~\eqref{eq33} and impose that its right hand side be orthogonal to $1$ and to $e^{i\theta}$, thereby obtaining
\begin{eqnarray}
\mathcal{R}^{(0)}=-2\pi\mbox{Re}\left[\!\left(\frac{\partial}{\partial X}+i\frac{\partial}{\partial Y}\right)\!A\right]\!,\quad
\mathcal{A}^{(0)}=\frac{1}{\pi}C^{(2)}[1,e^{i\theta}]_1 rA-\frac{1}{4\pi}\!\left(\frac{\partial}{\partial X}-i\frac{\partial}{\partial Y}\right)\! r.\label{eq38}
\end{eqnarray}

Then Eq.~\eqref{eq33} has the solution
\begin{eqnarray}
f^{(2)}(\theta,t,\textbf{X},T)=\!\left[\frac{A^2 C^{(2)}[e^{i\theta},e^{i\theta}]_2}{1-C^{(1)}[e^{i2\theta}]_2} - \frac{\left(\frac{\partial}{\partial X} - i\frac{\partial}{\partial Y}\right)\!A}{2(1-C^{(1)}[e^{i2\theta}]_2)} \right]\!e^{i2\theta}+\mbox{cc}. \label{eq39}
\end{eqnarray}
We now insert Eqs.~\eqref{eq28} and \eqref{eq39} into Eq.~\eqref{eq34}, and impose the solvability conditions to the resulting equation. We obtain
\begin{eqnarray}
&&\mathcal{R}^{(1)}=-\mbox{Re}\left[C^{(2)}[1,e^{i\theta}]_1\!\left(\frac{\partial}{\partial X}+i\frac{\partial}{\partial Y}\right)\! rA\right]\!,\nonumber\\
&&\mathcal{A}^{(1)}=[\Gamma(r)-\mu|A|^2]A+\delta\Delta_XA +\gamma_1\!\left(\frac{\partial}{\partial X}+i\frac{\partial}{\partial Y}\right)\!A^2
+\gamma_2\overline{A} \!\left(\frac{\partial}{\partial X}-i\frac{\partial}{\partial Y}\right)\!A\nonumber\\
&&+ \gamma_3A\!\left(\frac{\partial}{\partial X}-i\frac{\partial}{\partial Y}\right)\!\overline{A}
+\frac{\gamma_3}{8\pi^2} \!\left(\frac{\partial}{\partial X}-i\frac{\partial}{\partial Y}\right)\! r^2.\label{eq40}
\end{eqnarray}
Here
\begin{eqnarray}
&&\Gamma(r)=\eta_2Q_\eta+\!\left(C^{(3)}[1,1,e^{i\theta}]_1-2 (C^{(2)}[1,e^{i•\theta}]_1)^2\right)\!\frac{r^2}{4\pi^2},\quad Q_\eta\!=\!\frac{\partial}{\partial\eta}\!\left(\ln C^{(1)}[e^{i\theta}]_1\right)\big|_{\eta_c}\!, \nonumber\\
&&\delta=\frac{1+C^{(1)}[e^{i2\theta}]_2}{8(1-C^{(1)}[e^{i2\theta}]_2)},\quad\gamma_1=\frac{1}{4}C^{(2)}[1,e^{i\theta}]_1-\frac{C^{(2)}[e^{i\theta},e^{i\theta}]_2}{2(1-C^{(1)}[e^{i2\theta}]_2)},\quad 
\gamma_2=-\frac{C^{(2)}[e^{-i\theta},e^{i2\theta}]_1}{1-C^{(1)}[e^{i2\theta}]_2},\nonumber\\
&&\gamma_3=C^{(2)}[1,e^{i\theta}]_1,\quad \mu= \frac{2C^{(2)}[e^{i\theta},e^{i\theta}]_2C^{(2)}[e^{i2\theta},e^{-i\theta}]_1}{C^{(1)}[e^{i2\theta}]_2-1} - C^{(3)}[e^{i\theta},e^{i\theta},e^{-i\theta}]_1.   \label{eq41}
\end{eqnarray}
\end{widetext}

The coefficients listed in Eq.~\eqref{eq41} are calculated in Appendix \ref{ap:1}. In terms of the mean current density $\mathbf{w}$ defined as
\begin{equation}
A=\frac{w_x-iw_y}{2\pi}, \quad\mathbf{w}=(w_x,w_y),  \label{eq42}
\end{equation}
Eqs.\eqref{eq31}, \eqref{eq32}, \eqref{eq38} and \eqref{eq40} can be written as
\begin{eqnarray}
&&\frac{\partial r}{\partial T}+\nabla_X\!\cdot\!\left[\!\left(1+\frac{\epsilon\gamma_3r}{2\pi}\right)\!\mathbf{w}\right]\!=0,\label{eq43}\\
\frac{\partial \mathbf{w}}{\partial T}\!&\!=\!&\!-\frac{1}{2}\nabla_X\!\left[\!\left(1-\frac{\epsilon\gamma_3}{4\pi}r\right)\!r+\frac{\epsilon(2\gamma_1-\gamma_2-\gamma_3)}{2\pi}|\mathbf{w}|^2\!\right]\! \nonumber\\
&\!+\!& \!\epsilon\frac{2\gamma_1+\gamma_2-\gamma_3}{2\pi}(\mathbf{w}\cdot\nabla_X)\mathbf{w} \nonumber\\
&\!+\!& \!\epsilon\frac{2\gamma_1-\gamma_2+\gamma_3}{2\pi}\mathbf{w}(\nabla_X\!\cdot\mathbf{w}) +\epsilon\delta\nabla^2_X \mathbf{w}\!\nonumber\\
&\!+\!&\!\!\left[\frac{\gamma_3 r}{\pi}+\epsilon\!\left(\Gamma(r)\!-\!\frac{\mu |\mathbf{w}|^2}{4\pi^2}\right)\!\right]\! \mathbf{w}. \label{eq44}
\end{eqnarray}
For $\epsilon=0$, Eq.~\eqref{eq43} is the continuity equation for a density variable $r$ and a current density $\mathbf{w}$, which explains the name of the latter variable. The overall density of particles equals the average density $N/L^2$, which implies the following constraint for $r(\mathbf{X},T)$:
\begin{equation}
\int r(\mathbf{X},T)\, d\mathbf{X}=0.\label{eq45}
\end{equation}
\bigskip

\noindent {\em Remark 1.} All the parameters in Eqs.~\eqref{eq43}-\eqref{eq44} are real numbers. For $r=0$ and $\gamma_3= 0$, these equations are exactly equivalent to the amplitude equation (132) of Ref.~\cite{ihl16}. \bigskip

\noindent {\em Remark 2.} For $r\neq 0$, Eqs.~\eqref{eq43}-\eqref{eq44} are not equivalent to those in \cite{ihl16}. The differences are due to two inconsistencies in Ihle's approach \cite{ihl16}. Firstly, Ihle did not expand the eigenvalue $Q_1=C^{(1)}[e^{i\theta}]_1$ in powers of $\epsilon$. Instead, he introduced $(Q_1-1)\mathbf{w}$ as a linear term in the amplitude equation for $\mathbf{w}$ [cf. Eq.~\eqref{eq44}]. This is equivalent to setting $\gamma_3=0$ and $\Gamma(r)=Q_1-1$ in Eq.~\eqref{eq44}. Thus, Ihle's equations do not contain the quadratic term proportional to $r\mathbf{w}$ appearing in Eq.~\eqref{eq44}. Secondly, we have used a consistent perturbation scheme in the parameter $\epsilon$ whereas the Chapman-Enskog procedure of Ref.~\cite{ihl16} mixes different orders in $\epsilon$. For instance, the tensors $\mathbf{\Omega}_1$ and $\mathbf{\Omega}_3$ defined in Eqs.~(117)-(119) and (121) of Ref.~\cite{ihl16} are $O(\epsilon^2)$ whereas $\mathbf{\Omega}_j=O(\epsilon^3)$ for $j=2,4,5$. However, all these tensors enter in the equation for the current density, Eq.~(130) of Ref.~\cite{ihl16}, with equal footing. On the other hand, for uniform time-independent density, all terms in Ihle's Eq.~(132) are of order $\epsilon^3$ provided $\partial/\partial t=O(\epsilon^2)$, $\nabla=O(\epsilon)$, $\lambda-1= O(\epsilon^2)$ and $\mathbf{w}= O(\epsilon)$.\bigskip

\noindent {\em Remark 3.} The equations of motion for the average velocity in bird flocking proposed by Toner and Tu do not contain the quadratic term $r\mathbf{w}$ in Eq.~\eqref{eq44} \cite{ton95,ton05}. \bigskip

\noindent {\em Remark 4.} What happens if the noise scales differently with $\epsilon$? We have chosen a parabolic scaling, $\eta=\eta_c+\epsilon^2\eta_2$. The only case that affects differently the outcome in Eqs.~\eqref{eq43}-\eqref{eq44} is $\eta=\eta_c+\epsilon\eta_1$. Then $\mathcal{A}^{(0)}$ has to include an additional term $\eta_1Q_\eta$ in Eq.~\eqref{eq38}. This means we have to write $r^*=r+\pi\eta_1Q_\eta/C^{(2)}[1,e^{i\theta}]_1$ instead of $r$ in Eqs.~\eqref{eq43}-\eqref{eq44}. In Eq.~\eqref{eq44}, we also have to replace  the term $Q_{\eta\eta}\eta_1^2/2$ (where $Q_{\eta\eta}=\partial^2Q_1/\partial\eta^2|_{\eta=\eta_c}$) instead of $Q_\eta\eta_2$ in $\Gamma(r)$ given by Eq.~\eqref{eq41}. The resulting $\Gamma(r^*)$ is negative (at least for the numerical values $M=7$, $\rho_0=5$ used in our simulations). We will not obtain a consistent stationary space independent solution of Eqs.~\eqref{eq43}-\eqref{eq44} unless $r^*=O(\epsilon)$, which will take us back to the parabolic scaling of the noise.

\section{Space-independent amplitude equation}\label{sec:6}
\begin{figure}[h]
\begin{center}
\includegraphics[width=9cm]{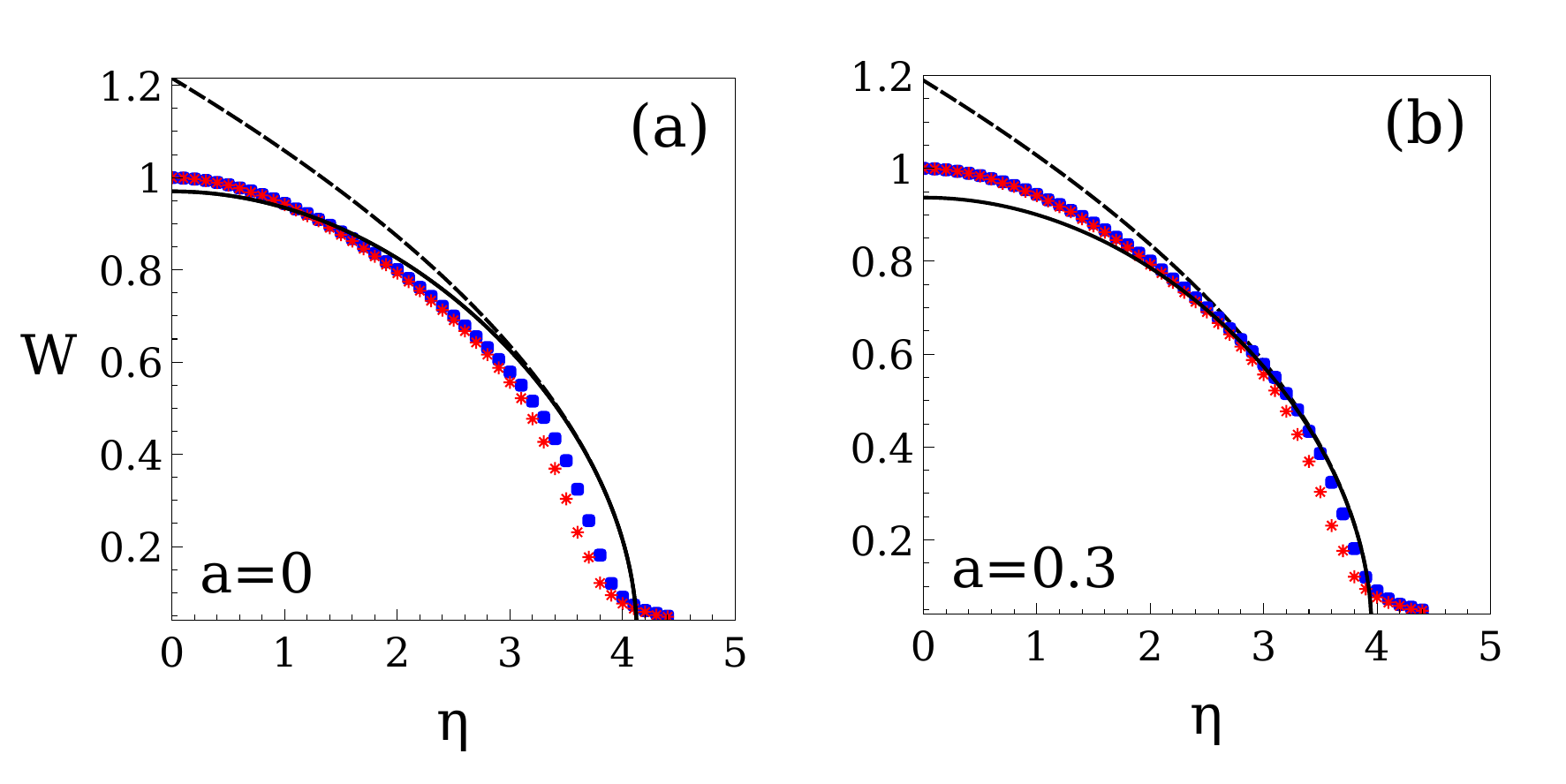}
\end{center}
\caption{(a) Polarization $W$ versus $\eta$ for $M=7$, $N=1000$, and $\rho_0=10$ (blue squares) or $\rho_0=5$ (red asterisks). Dashed and solid lines correspond to Eqs.~\eqref{eq49} and \eqref{eq50}, respectively. (b) Same graph with the critical noise shifted according to Eq.~\eqref{eq69} with $a=(\epsilon/\rho_0)\sqrt{\langle\tilde{r}^2\rangle/6}=0.3$.
 \label{fig2}}
\end{figure}

For space independent $A$ and $r=0$, Eq.~\eqref{eq32} is the typical pitchfork amplitude equation
\begin{equation}
\frac{\partial A}{\partial(\epsilon^2t)}=(\eta_2Q_\eta-\mu |A|^2)A.  \label{eq46}
\end{equation}
Note that time scales as the square of space (diffusive scaling). As $\mu>0$ and $Q_\eta<0$, the stationary solution 
\begin{equation}
A=\sqrt{\frac{\eta_2Q_\eta}{\mu}}\, e^{i\Upsilon}, \quad\Upsilon\in\mathbb{R},\label{eq47}
\end{equation}
is stable and it exists for $\eta<\eta_c$. In this region, the uniform distribution $f_0$ is unstable because $Q_1>1$. Thus the transition from incoherence to order is a supercritical pitchfork bifurcation, as depicted in Fig.~\ref{fig2}.

Fig.~\ref{fig2} represents the polarization given by the modulus of the complex parameter \eqref{eq6}. According to Eq.~\eqref{eq28}, the order parameter is
\begin{eqnarray}
Z= \frac{1}{N}\sum_{j=1}^N e^{i\theta_j}=\frac{1}{\rho_0L^2}\int e^{i\theta}f(\theta,\mathbf{x},t)\, d\theta d\mathbf{x}\nonumber\\
\sim\frac{2\pi\epsilon}{\rho_0\epsilon^2L^2}\int\overline{A(\mathbf{X},T)}\, d\mathbf{X}.\label{eq48}
\end{eqnarray}
For the uniform solution given by Eq.~\eqref{eq47}, the order parameter is 
\begin{eqnarray}
Z_0\sim\frac{2\pi}{\rho_0} \sqrt{\frac{(\eta-\eta_c)Q_\eta}{\mu}}\, e^{-i\Upsilon}.  \label{eq49}
\end{eqnarray}
Near $\eta_c$, we have $Q_1\sim 1+ Q_\eta(\eta-\eta_c)$. Then we can replace $Q_\eta(\eta-\eta_c)$ by $Q_1-1$ in Eq.~\eqref{eq49}, thereby getting a formula that holds for larger values of $|\eta-\eta_c|$ (cf.~Fig.~\ref{fig2}):
\begin{eqnarray}
Z_0\sim\frac{2\pi}{\rho_0} \sqrt{\frac{Q_1-1}{\mu}}\, e^{-i\Upsilon}.  \label{eq50}
\end{eqnarray}
We observe that the values obtained by direct simulations of the VM tend to the predicted solution as we increase the density $\rho_0$ from 5 to 10 in Fig.~\ref{fig2}(a). The change in convexity of the curve near $\eta_c$ is due to finite size effects. Moreover, Fig.~\ref{fig2}(b) shows that the critical noise found in the simulations of the VM is shifted to $\eta_c-a^2$ with $a=0.3$ for $\rho_0=10$.

\section{Space-dependent amplitude equations}\label{sec:7}
For nonzero $r$ and space dependent $A$, we need to consider the space-dependent Eqs.~\eqref{eq43}-\eqref{eq44}. The leading order equations for $\epsilon=0$ are 
\begin{eqnarray}
&&\frac{\partial r}{\partial T}+\nabla_X\!\cdot\!\mathbf{w}=0,\label{eq51}\\
&&\frac{\partial \mathbf{w}}{\partial T}=-\frac{1}{2}\nabla_X r+\frac{\gamma_3}{\pi}r \mathbf{w}, \label{eq52}
\end{eqnarray}
In these equations, space and time scale in the same way as $\mathbf{X}=\epsilon\mathbf{x}$ and $T=\epsilon t$ (hyperbolic or convective scaling). The order parameter \eqref{eq48} can now be used to define a vector order parameter, $\mathbf{Z}=$(Re$Z$,Im$Z$):
\begin{eqnarray}
\mathbf{Z}\sim \frac{\epsilon}{\rho_0L^2}\int \mathbf{w}(\mathbf{X},T)\, dx\, dy,  \label{eq53}
\end{eqnarray}
because $\overline{A(\mathbf{X},T)}=[w_x(\mathbf{X},T)+iw_y(\mathbf{X},T)]/(2\pi)$, according to Eq.~\eqref{eq42}. Ignoring the nonlinear term in Eq.~\eqref{eq52} (small initial data), we eliminate the current density and obtain the linear wave equation with velocity $1/\sqrt{2}$. Then $\nabla_X\!\cdot\!\mathbf{w}$ obeys the same wave equation and $\nabla_X\!\times\!\mathbf{w}$ is independent of time. The overall constraint \eqref{eq45} holds for all time provided it does so initially. Space independent solutions of this system produce a current density that increases with time if $r>0$ and decreases if $r<0$. It seems that shock waves may form after a finite time.

\subsection{1D equations for hyperbolic scaling} In 1D, Eqs.~\eqref{eq51}-\eqref{eq52} can be written in the form of conservation laws as
\begin{eqnarray}
&&\frac{\partial\zeta_{\pm}}{\partial T}\pm\frac{1}{\sqrt{2}}\frac{\partial\zeta_{\pm}}{\partial X}=\pm\gamma_3\frac{\zeta_+^2-\zeta_-^2}{4\pi}, \label{eq54}\\
&&\zeta_{\pm}=r\pm\sqrt{2}\,w. \label{eq55}
\end{eqnarray}
Eqs.~\eqref{eq54}-\eqref{eq55} with $u=-\zeta_+$ and $v=-\zeta_-$ are the Carleman model \cite{god71}, but with negative values of $u$ and $v$. For sufficiently small values of $\rho_0$ and appropriate boundary conditions (not periodic ones), it is known that the solutions of this model present chaotic regimes in these circumstances \cite{ari10,rad17}.

\subsection{Linearized 2D equations for convective scaling} 
Figs.~\ref{fig2} show that, for the long times employed in direct simulations of the VM, there is agreement between the simulations and the uniform solution \eqref{eq47} of Eqs.~\eqref{eq43}-\eqref{eq44},
\begin{equation}
r_0\!=0,\,\mathbf{w}_0\!=2\pi\sqrt{\frac{\eta_2Q_\eta}{\mu}}\mathbf{e}_\Upsilon,\, \mathbf{e}_\Upsilon\!=(\cos\Upsilon,-\sin\Upsilon). \label{eq56}
\end{equation}
However, very close to $\eta_c$, the separation between the hyperbolic and parabolic time scales has appreciable effects. To uncover them, we linearize Eqs.~\eqref{eq51}-\eqref{eq52} about Eq.~\eqref{eq56}, thereby obtaining
\begin{eqnarray}
&&\frac{\partial\tilde{r}}{\partial T}+\nabla_X\!\cdot\!\mathbf{\tilde{w}}=0,\label{eq57}\\
&&\frac{\partial \mathbf{\tilde{w}}}{\partial T}=-\frac{1}{2}\nabla_X\tilde{r}+\frac{\gamma_3}{\pi}\mathbf{w}_0\tilde{r}. \label{eq58}
\end{eqnarray}
By differentiating Eq.~\eqref{eq57} and eliminating $\mathbf{\tilde{w}}$ by means of Eq.~\eqref{eq58}, we find the wave equation:
\begin{equation}
\frac{\partial^2\tilde{r}}{\partial T^2}=\frac{1}{2}\nabla_X^2\tilde{r} -\frac{\gamma_3}{\pi}\, \mathbf{w}_0\cdot\nabla_X\tilde{r}. \label{eq59}
\end{equation}
For periodic boundary conditions, we can solve this equation by writing $\tilde{r}(\mathbf{X},T)$ as a Fourier series on the square box of size $L$. Then we can find $\mathbf{\tilde{w}}(\mathbf{X},T)$ from Eq.~\eqref{eq58}. However, the gradient term produces a combination of factors exponentially increasing with time and factors exponentially decreasing with time. It is then hard to predict the long time behavior of the solutions. We can obtain an equivalent formulation by using the change of variable
\begin{equation}
\tilde{r} = e^{\gamma_3\mathbf{w}_0\cdot\mathbf{X}/\pi} R, \label{eq60}
\end{equation}
to eliminate the gradient term in Eq.~\eqref{eq59}, thereby producing the Klein-Gordon equation:
\begin{equation}
\frac{\partial^2R}{\partial T^2}=\frac{1}{2}\nabla_X^2R -\frac{\gamma_3^2|\mathbf{w}_0|^2}{2\pi^2}\, R.\label{eq61}
\end{equation}
For periodic boundary conditions, $R(\mathbf{X},T)=\sum_{n,m}R_{n,m}(T)e^{i\mathbf{k}_{n,m}\cdot\mathbf{X}}$, and the coefficients $R_{n,m}(T)$ solve the equation of a linear oscillator with frequency
\begin{equation}
\omega_{n,m}=\sqrt{\frac{1}{2}|\mathbf{k}_{n,m}|^2 +\frac{\gamma_3^2|\mathbf{w}_0|^2}{2\pi^2}}, \,\,\mathbf{k}_{n,m}=\frac{2\pi}{\epsilon L}(n,m). \label{eq62}
\end{equation}
Note that the frequency $\omega_{n,m}$ mixes frequencies corresponding to the acoustic velocity $1/\sqrt{2}$ [cf. $|\mathbf{w}_0|=0$ in Eq.~\eqref{eq62}] with the fundamental mode of frequency $\gamma_3|\mathbf{w}_0|/(\sqrt{2}\pi)$ corresponding to $n=m=0$. 

We now solve the equation for $R_{n,m}(T)$ and then reconstruct $\tilde{r}(\mathbf{X},T)$ and $\mathbf{\tilde{w}}(\mathbf{X},T)$ from Eqs~\eqref{eq51}-\eqref{eq52} and \eqref{eq60}. The results are
\begin{widetext}
\begin{eqnarray}
&&\tilde{r}(\mathbf{X},T)=\sum_{n=-\infty}^\infty\sum_{m=-\infty}^\infty \! \left[R_{n,m}(0)\cos(\omega_{n,m}T)+\frac{\dot{R}_{n,m}(0)}{\omega_{n,m}}\sin(\omega_{n,m}T)\right] e^{(i\mathbf{k}_{n,m}+\frac{\gamma_3\mathbf{w}_0}{\pi})\cdot\mathbf{X}}, \label{eq63}\\
&&\mathbf{\tilde{w}}(\mathbf{X},T) =\frac{1}{2}\sum_{n=-\infty}^\infty\sum_{m=-\infty}^\infty\! \left(i\mathbf{k}_{n,m}-\frac{\gamma_3\mathbf{w}_0}{\pi}\right)\!\frac{e^{(i\mathbf{k}_{n,m}+\frac{\gamma_3\mathbf{w}_0}{\pi})\cdot\mathbf{X}}}{\omega_{n,m}}\! \left[R_{n,m}(0) \sin(\omega_{n,m}T)-\frac{\dot{R}_{n,m}(0)}{\omega_{n,m}}\cos(\omega_{n,m}T)\right] \nonumber\\
&&\quad\quad\quad\quad +\sum_{n=-\infty}^\infty\sum_{m=-\infty}^\infty\mathbf{C}_{m,n}\, e^{i\mathbf{k}_{n,m}\cdot\mathbf{X}}, \label{eq64}\\
&& R_{n,m}(0)\!= \!\frac{1}{\epsilon^2L^2}\!\int_0^{\epsilon L}\!\!\!\int_0^{\epsilon L}\!\!\!\! e^{-(i\mathbf{k}_{n,m}+\frac{\gamma_3\mathbf{w}_0}{\pi})\cdot\mathbf{X}}\, \tilde{r}(\mathbf{X},0)\, d\mathbf{X}\!=\!\! \sum_{l=-\infty}^\infty\sum_{j=-\infty}^\infty\! \frac{\tilde{r}_{l,j}(0)\, (e^{\frac{\epsilon\gamma_3 w_{0x}L}{\pi}}-1)(e^{\frac{\epsilon\gamma_3 w_{0y}L}{\pi}}-1)}{\!\left[\frac{\epsilon\gamma_3 w_{0x}L}{\pi}+i2\pi(l-n)\right]\!\!\left[\frac{\epsilon\gamma_3 w_{0y}L}{\pi}+i2\pi(j-m)\right]\!},\quad \label{eq65}\\
&& \dot{R}_{n,m}(0)= - \frac{1}{\epsilon^2L^2}\int_0^{\epsilon L}\int_0^{\epsilon L} e^{-(i\mathbf{k}_{n,m}+\frac{\gamma_3\mathbf{w}_0}{\pi})\cdot\mathbf{X}} \!\left(i\mathbf{k}_{n,m}+\frac{\gamma_3\mathbf{w}_0}{\pi}\right)\!\cdot\mathbf{\tilde{w}}(\mathbf{X},0)\, dX\, dY, \label{eq66}\\
&& \mathbf{C}_{n,m}= \mathbf{\tilde{w}}_{n,m}(0)+\frac{1}{2}\sum_{l=-\infty}^\infty\sum_{j=-\infty}^\infty \left(i\mathbf{k}_{l,j}-\frac{\gamma_3\mathbf{w}_0}{\pi}\right) \!\frac{e^{\frac{\epsilon\gamma_3 w_{0x}L}{\pi}}-1}{\frac{\epsilon\gamma_3 w_{0x}L}{\pi}+i2\pi(l-n)}\, \frac{e^{\frac{\epsilon\gamma_3 w_{0y}L}{\pi}}-1}{\frac{\epsilon\gamma_3 w_{0y}L}{\pi}+i2\pi(j-m)}\,\frac{\dot{R}_{l,j}(0)}{\omega_{l,j}^2}. \label{eq67}
\end{eqnarray}
\end{widetext}
For $\mathbf{w}(\mathbf{X},T)=\mathbf{w}_0+\mathbf{\tilde{w}}(\mathbf{X},T)$, the order parameter \eqref{eq53} becomes
\begin{eqnarray}
&&\mathbf{Z}\sim\frac{\epsilon}{\rho_0}\mathbf{w}_{0}+ \frac{\epsilon}{\rho_0\epsilon^2L^2} \int_0^{\epsilon L}\! \int_0^{\epsilon L} \mathbf{\tilde{w}}(\mathbf{X},T)\, dX\,dY\quad\nonumber\\
&&\quad= \frac{\epsilon}{\rho_0}\mathbf{w}_{0}+\frac{\epsilon}{\rho_0}\mathbf{\tilde{w}}_{0,0}(T). 
\label{eq68}
\end{eqnarray}

\subsection{Shift in the critical noise} 
The oscillating density disturbance $\tilde{r}(\mathbf{X},T)$ will produce a nonzero value of the average of $r^2=\tilde{r}^2$. Averaging Eq.~\eqref{eq44} over space and time, and assuming that the average of a product is the product of averages, we have $\langle r\mathbf{w}\rangle\approx\langle r\rangle\,\langle\mathbf{w}\rangle=0$ and $\langle r^2\mathbf{w}\rangle\approx\langle r^2\rangle\,\langle\mathbf{w}\rangle$. Then the time-independent and space-averaged part of the term $r^2$ in $\Gamma(r)$ gives a contribution to $|\mathbf{w}_0|=2\pi\sqrt{\Gamma(r)/\mu}$, which yields the first term in Eq.~\eqref{eq68}: 
\begin{eqnarray}
\!\frac{\epsilon\mathbf{w}_0}{\rho_0}\!&\sim&\!\frac{2\pi}{\rho_0} \sqrt{\frac{Q_\eta (\eta-\eta_c) -\frac{\epsilon^2\langle\tilde{r}^2(\mathbf{X},T)\rangle}{6\rho_0^2}}{\mu}}\,\mathbf{e}_\Upsilon \nonumber\\
\!&\sim&\! \frac{2\pi}{\rho_0} \sqrt{\frac{Q_1-1-\frac{\epsilon^2\langle\tilde{r}^2(\mathbf{X},T)\rangle}{6\rho_0^2}}{\mu}}\,\mathbf{e}_\Upsilon. \label{eq69}
\end{eqnarray}
Here $\tilde{r}$ is given by Eq.~\eqref{eq63} and, inserted into the average over time and space in Eq.~\eqref{eq69}, contributes to shift the bifurcation point to a smaller noise value $\eta_c^*$. Fig.~\ref{fig2} shows that simulation data are consistent with $\epsilon\sqrt{\langle\tilde{r}^2(\mathbf{X},T)\rangle/6}= a\rho_0$, with $a\approx 0.3$. Then $\eta_c^*\approx 3.95$. One first correction consists of using a more accurate formula instead of Eq.~\eqref{eq23} for finite values of $M$. Using the same procedure as explained in Ref.~\cite{ihl16} but keeping more terms in the expansions, the critical condition $Q_1=1$ for the noise becomes
\begin{eqnarray}
 \frac{\sqrt{\pi M}}{\eta}\left(1-\frac{1}{8M}-\frac{7}{128 M^2}-\frac{5}{128M^3}\right)\sin\frac{\eta}{2}=1. \label{eq70}  
\end{eqnarray}
See Appendix \ref{ap:4}. For $M=7$, we obtain $\eta_c=4.09$ ($a=0.26$) instead of the theoretical value $\eta_c=4.13$, as in Fig.~\ref{fig2}(a). Taking into account that the second term in Eq.~\eqref{eq68} is obtained from linearization about the first term (and is therefore assumed to be small compared with it), the polarization becomes $|\mathbf{Z}|\sim(\epsilon/\rho_0)\,|\mathbf{w}_0\!+\!\mathbf{\tilde{w}}_{0,0}|\sim(\epsilon/\rho_0)(|\mathbf{w}_0|+\mathbf{w}_0\cdot\mathbf{\tilde{w}}_{0,0}/|\mathbf{w}_0|)$, i.e., 
\begin{eqnarray}
|\mathbf{Z}|\sim \frac{2\pi}{\rho_0} \sqrt{\frac{Q_1-1-\frac{\epsilon^2\langle\tilde{r}^2(\mathbf{X},T)\rangle}{6\rho_0^2}}{\mu}}+\frac{\epsilon}{\rho_0}\mathbf{e}_\Upsilon\!\cdot\!\mathbf{\tilde{w}}_{0,0}(T). \label{eq71}
\end{eqnarray}
In Appendix \ref{ap:4}, we estimate a value for the shift after some uncontrolled approximations that take advantage of the linearized equations \eqref{eq57}-\eqref{eq59}. The result $a\approx 0.01$ is a small improvement that agrees better with the numerically estimated shift of the bifurcation value. Better agreement should be achieved by numerically solving  the full nonlinear equations \eqref{eq43}-\eqref{eq44} and finding a more precise value of the time and space average $\langle r^2\rangle$ in the formula for $|\mathbf{w}_0|$. The shift in critical noise was noticed earlier in Ref.~\cite{cho12} but no explanation thereof was given there.

\begin{figure}[h]
\begin{center}
\includegraphics[width=8.5cm]{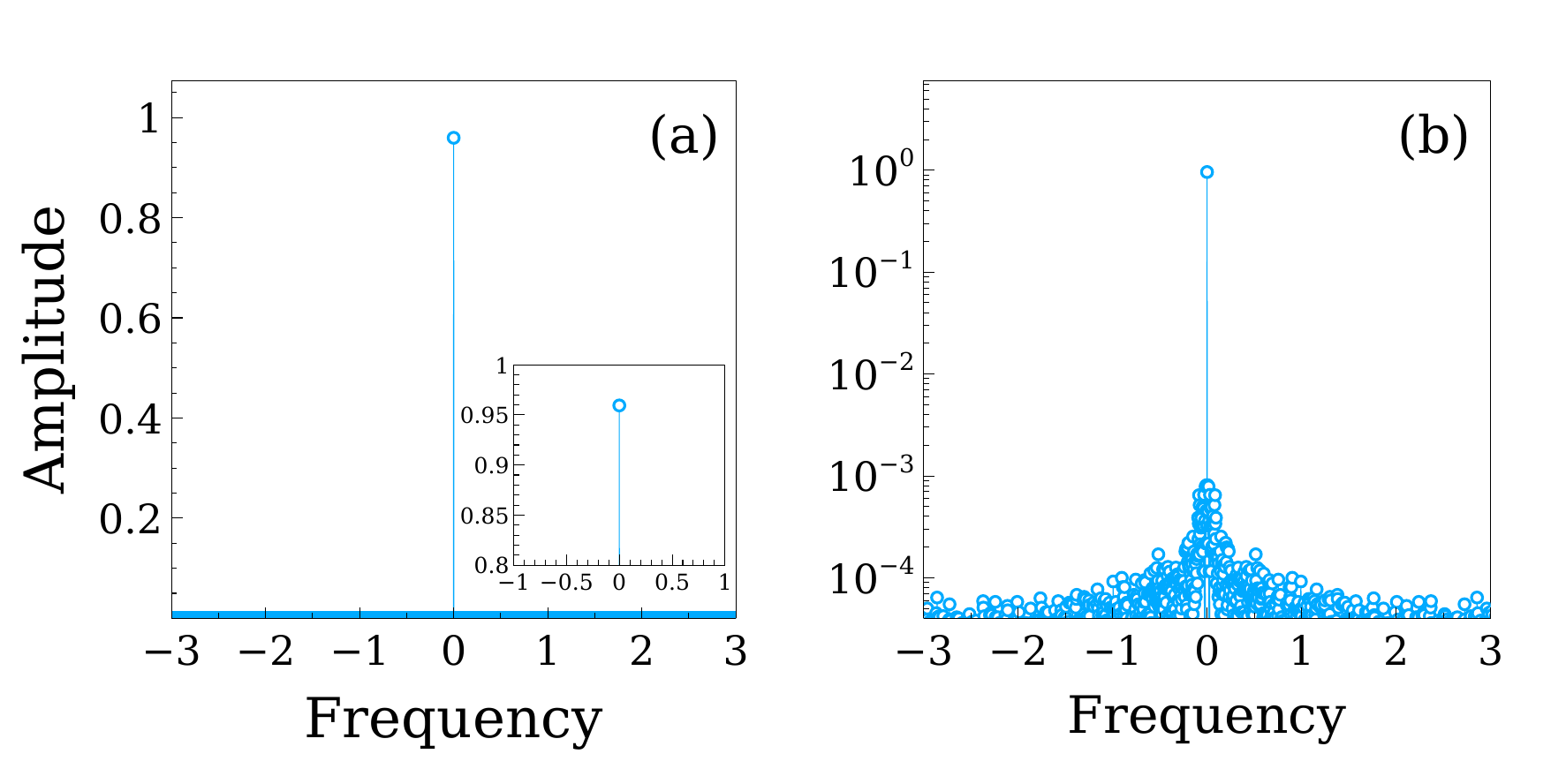}\\
\includegraphics[width=8.5cm]{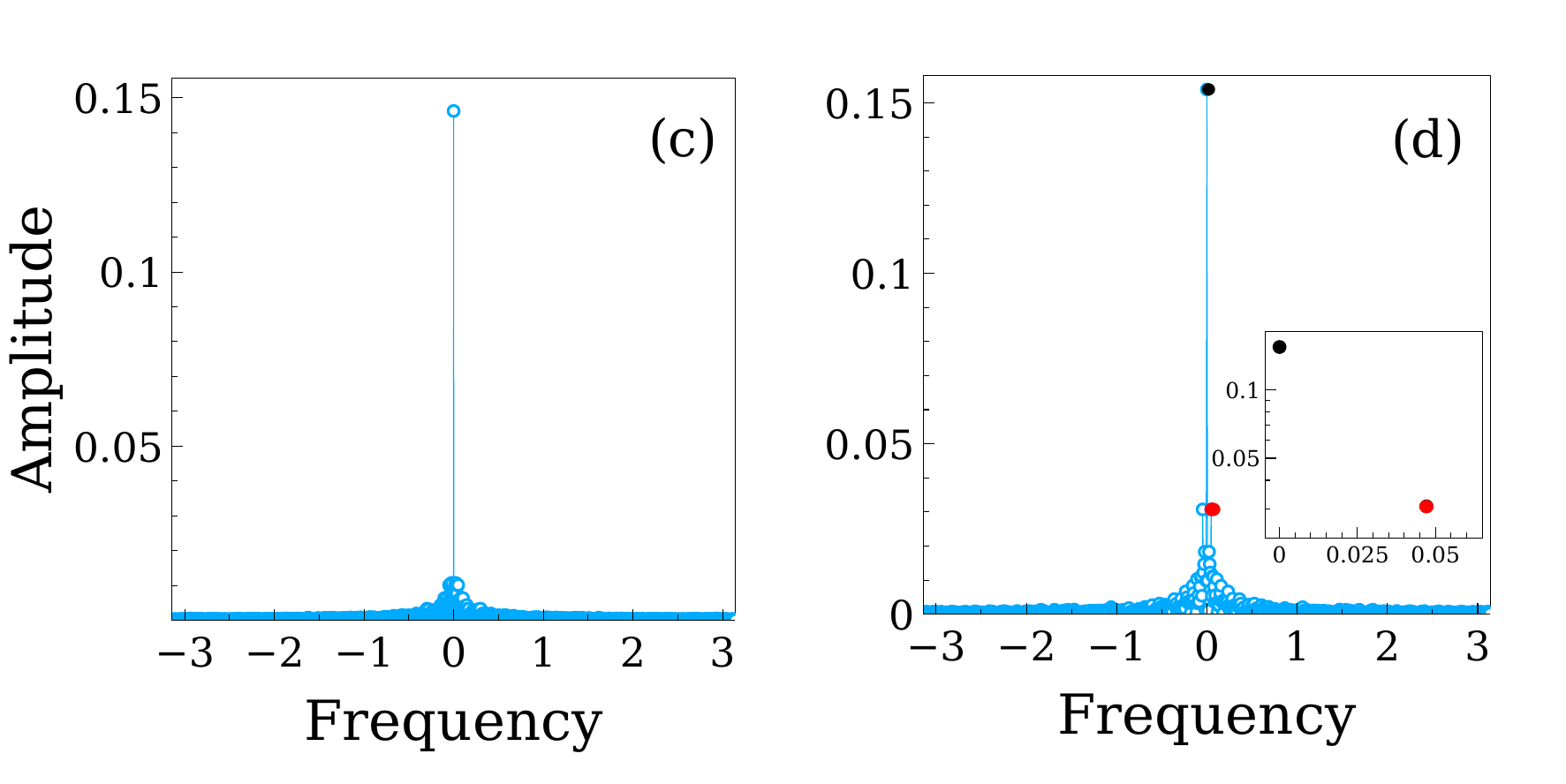}
\end{center}
\caption{Discrete Fourier transform of the polarization $W(t)$ obtained from single simulations of the VM (without ensemble averaging) for: (a), (b) and (c) Vicsek alignment rule of Eq.~\eqref{eq2}; (d)  alignment rule of Eq.~\eqref{eq74}, which includes harmonic forcing of frequency $\omega=\Omega_Z\approx 0.05$ with $h=5$. Parameter values are $M=7$, $N=1000$, $\rho_0=5$, and (a)-(b) $\eta=0.8$, (c)-(d) $\eta=3.7$. Panel (b) is the same as Panel (a) in logarithmic scale. The inset of Panel (a) shows that, for this single simulation, the amplitude of the zero-frequency mode coincides with the ensemble averaged value of $W$ in Fig.~\ref{fig2}. In Panel (d), filled circles indicate the zero (black dot) and resonant (red dot) frequencies, both peaks are highlighted in the Inset. \label{fig3}}
\end{figure}

\subsection{Oscillatory correction to the polarization: resonances in the Vicsek model} 
Eq.~\eqref{eq71} contains a bounded oscillatory function of $T$ [cf.~Eq.~\eqref{eq64}]. Thus the polarization is a function of the time $t$ whose lowest angular frequency is $\Omega_Z$: 
\begin{eqnarray}
\Omega_Z\sim 2\gamma_3\sqrt{\frac{Q_1-1-\frac{\epsilon^2\langle\tilde{r}^2(\mathbf{X},T)\rangle}{6\rho_0^2}}{2\mu}}\sim\frac{W}{2\sqrt{2}}. \label{eq72}
\end{eqnarray}
Here $W$ is given by Eq.~\eqref{eq69}. The other frequencies given by Eq.~\eqref{eq62} are now
\begin{equation}
\epsilon\,\omega_{n,m}\sim \sqrt{\Omega_Z^2+\frac{2\pi^2}{L^2}(n^2+m^2)}\,. \label{eq73}
\end{equation}

How can we confirm the existence of these oscillation frequencies? One possibility is to modify the alignment rule from Eq.~\eqref{eq2} to
\begin{eqnarray}
\theta_i(t+\tau)\!=\!\mbox{Arg}\!\!\left(\sum_{|\mathbf{x}_j\!-\!\mathbf{x}_i|<R_0}\!e^{i\theta_j(t)}\!\right)\!\!+\!\xi_i(t)\!+\! h\cos(\omega t),\,\, \label{eq74}
\end{eqnarray}
move the forcing frequency until it resonates with one of the frequencies of Eq.~\eqref{eq73}, and simulate the resulting forced VM. However, we need to explore a region of $\eta$ sufficiently close to $\eta_c$. Otherwise, the parameter $\epsilon=\sqrt{(\eta-\eta_c)/\eta_2}$ is so large that there is no separation between the hyperbolic and parabolic scalings. In this later case, say for $\eta=0.8$, the discrete Fourier transform of the polarization $W(t)$, shown in Fig.~\ref{fig3}(a), has a single peak at zero frequency and a small, seemingly flat, background [the amplitudes of the transform at nonzero frequency are all smaller than $0.001$, cf. Fig.~\ref{fig3}(b)]. For the transform depicted in Figs.~\ref{fig3}(a) and \ref{fig3}(b), the peak height at zero frequency coincides with the value displayed in Fig.~\ref{fig2}, which has been obtained as an ensemble average over many realizations of the stochastic process given by the VM. As $\eta$ increases towards $\eta_c$, one simulation of the VM shows that $W(t)$ still has a large peak at zero frequency, but there is a small mound about it [cf.~Fig.~\ref{fig3}(c)]. If we repeat the simulations of the VM with the modified rule Eq.~\eqref{eq74}, Fig.~\ref{fig3}(d) shows that the mound is higher and that there is a small resonant peak at $\omega= \Omega_Z$. This effect is very small because, for $\eta$ close to $\eta_c$, the polarization and, consequently, the frequency given by Eq.~\eqref{eq72}, are very small, and the alignment noise is large.  

\begin{figure}[h]
\begin{center}
\includegraphics[width=5cm]{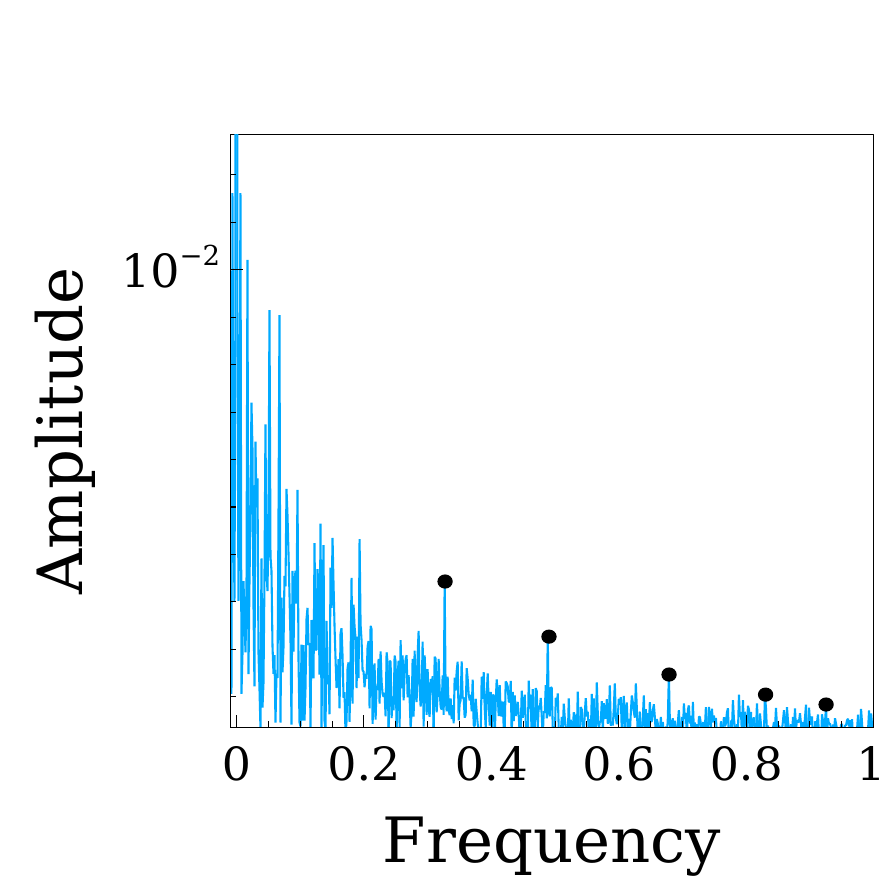}
\end{center}
\caption{Same as Fig.~\ref{fig3} but with the forcing frequency of Eq.~\eqref{eq73} with $n=m=2$ and $h=5$.  \label{fig4}}
\end{figure}

We can also excite higher frequencies by setting the forcing frequency to be one of those in Eq.~\eqref{eq73}. For instance, if we set $\omega$ in Eq.~\eqref{eq74} equal to the frequency given by Eq.~\eqref{eq73} with $n=m=2$, the nonlinearity of the amplitude equations excites several nearby frequencies. Thus, the peaks appearing in Fig.~\ref{fig4} are close to frequencies of the modes (i) $(n,m) = (1,0),\, (0,1)$, (ii) $(1,1)$, (iii) $(2,0),\, (0,2)$, (iv) $(2,2)$, and (v) $(3,0),\,(0,3)$, given by Eq.~\eqref{eq73}. 

\section{Final Remarks}\label{sec:8}
We have studied flocking in the standard Vicsek model by analyzing the bifurcation of the uniform distribution function to solutions of the associated kinetic equation that have nonzero polarization. Our linear stability analysis of the uniform distribution is limited to small wave numbers. Within this constraint, linear instability first occurs at a real eigenvalue with zero wave number. The picture of flocking that emerges from the bifurcation analysis is intricate. 

We have shown that the amplitude equations for the bifurcating modes near the critical value $\eta_c$ of the noise $\eta$ are equivalent to coupled equations for a disturbance of the number density and a current density. The equation for the density disturbance is a continuity equation whereas the equation for the current density contains two different scalings: (i) a hyperbolic scaling in which both time and space scale as $|\eta-\eta_c|^{-1/2}$; and (ii) a parabolic scaling in which time scales as $|\eta-\eta_c|^{-1}$, therefore, as space squared. 

Space-independent solutions of the amplitude equations obey the usual equation for a supercritical pitchfork bifurcation on the longer parabolic time scale. Stable stationary solutions of this equation produce a polarization proportional to $|\eta-\eta_c|^{1/2}$ as $\eta\to\eta_c$, which is depicted in Fig.~\ref{fig2}. Compared to direct simulations of the VM, ensemble averages of the polarization are similar to the bifurcation predictions, except for a shift of the critical noise to smaller values and a round off very close to $\eta_c$. 

Space-dependent patterns near the flocking bifurcation satisfy the full amplitude equations. On the longer parabolic time scale, the solutions of the latter should be close to the space-independent stationary solutions. Then we can study the linearization of the amplitude equations about such solutions. The leading order approximation of the linearized equations on the hyperbolic scaling is equivalent to a Klein-Gordon equation whose solutions for periodic boundary conditions contain infinitely many frequencies of oscillation. Thus, the emerging picture of the flocking bifurcation in the VM is that of an almost uniform polarization with small superimposed Klein-Gordon oscillations that occur on the faster hyperbolic scaling. These oscillations produce a nonzero average of the square of the density disturbance resulting in a shift in the critical noise, which may explain the observed one in direct simulations of the VM. 

To confirm this picture of flocking, we have modified the Vicsek alignment rule by adding a harmonic forcing term to the average alignment direction and to the noise. We have then simulated the resulting model looking for resonances between the forcing frequency and one of the Klein-Gordon frequencies. The discrete Fourier transform of the resulting time-dependent polarization shows a large peak at zero frequency (with amplitude equal to that of the stable stationary solution) and much smaller peaks at other frequencies. For large $|\eta-\eta_c|$, the parabolic and hyperbolic scalings are no longer separated and dissipation effects move the polarization to its stationary value. Close to the critical noise, both scalings are separated and the resonant peak emerges from the background in the discrete Fourier transform obtained from a single direct simulation of the VM. This effect is masked by the large size of the noise near its critical value but it is clearly observable in single simulations of the VM, as shown in Fig.~\ref{fig3}(d).

One caveat should be indicated here. If the box size is sufficiently large, the flocking transition is discontinuous and band-like patterns have been observed in numerical simulations of the VM \cite{gre04}. Our bifurcation theory for a mode with zero wave number is not applicable to these large box sizes. However, a future extension of our linear stability analysis of the uniform distribution to larger wave numbers may describe the resulting flocking transition, at least near the critical box size at which flocking becomes a discontinuous bifurcation.

\acknowledgements
We thank Antonio Lasanta for useful comments and for bringing the kinetic theory work of Ihle and collaborators to our attention. This work has been supported by the Ministerio de Econom\'\i a y Competitividad grants MTM2014-56948-C2-2-P and MTM2017-84446-C2-2-R. LLB thanks Russel Caflisch for hospitality during a sabbatical stay at the Courant Institute and acknowledges support of the Ministerio de Ciencia, Innovaci\'on y Universidades ``Salvador de Madariaga'' grant PRX18/00166.

\appendix 
\section{Eigenvalues for $|\mathbf{K}|=0$ in the limit as $n\to\infty$}\label{ap:2}
The eigenvalues $Q_j$ for zero wave number are given by Eqs.~\eqref{eq21}-\eqref{eq22}. To find them, we need to calculate integrals of the form
\begin{eqnarray}
\mathcal{J}(n,j)=\int_{[-\pi,\pi]^n}e^{ij(\theta_1-\Phi_1)}\prod_{l=1}^n \frac{d\theta_l}{2\pi}, \label{a2_1}
\end{eqnarray}
in the limit as $n\to\infty$. We have
\begin{eqnarray}
e^{ij(\theta_1-\Phi_1)}=\left(\frac{1+L e^{i(\theta_1-\beta)}}{\sqrt{(1+L e^{i(\theta_1-\beta)})(1+L e^{i(\beta-\theta_1)})}}\right)^j\!\nonumber\\
=\left(\frac{1+L e^{i(\theta_1-\beta)}}{1+L e^{i(\beta-\theta_1)}}\right)^{j/2}\!, \label{a2_2}
\end{eqnarray}
where
\begin{eqnarray}
L e^{i\beta}=\frac{\sum_{l=2}^n e^{i\theta_l}}{|\sum_{l=2}^ne^{i\theta_l}|}.\label{a2_3}
\end{eqnarray}
In the limit as $n\to\infty$, the central limit theorem implies that we can replace $n-1$ integrals (with $n-1\sim n$) in $\mathcal{J}(n,j)$ by \cite{ihl16}
\begin{widetext}
\begin{eqnarray}
\mathcal{J}(n,j)=\frac{1}{2\pi}\int_0^\infty\int_{-\pi}^\pi\int_{-\pi}^\pi \frac{L e^{-L^2/n}}{\pi n}\!\left(\frac{1+L e^{i(\theta_1-\beta)}}{1+L e^{i(\beta-\theta_1)}}\right)^{j/2}\!dL\, d\theta_1 d\beta=\int_0^\infty\int_{-\pi}^\pi \frac{L e^{-L^2/n}}{\pi n}\!\left(\frac{1+L e^{i\theta}}{1+L e^{-i\theta}}\right)^{j/2}\!dL\, d\theta. \label{a2_4}
\end{eqnarray}
\end{widetext}
We find different approximations for odd and even $j$. For odd $j$, we split the $L$-integral in sub-integrals over $(0,\Lambda)$ and $(\Lambda,\infty)$, with $\Lambda\gg 1$ fixed in the limit as $n\to\infty$. We can approximate $1/n=0$ in the first sub-integral and expand the fraction in powers of $1/L$ in the second sub-integral. The result is
\begin{eqnarray}
&&\mathcal{J}(n,j)=\frac{1}{n}\int_0^\Lambda\int_{-\pi}^\pi \frac{L}{\pi}\!\left(\frac{1+L e^{i\theta}}{1+L e^{-i\theta}}\right)^{j/2}\!dL\, d\theta\nonumber\\
&&+\int_\Lambda^\infty\int_{-\pi}^\pi \frac{L e^{ij\theta-L^2/n}}{\pi n}\!\left\{1-\frac{ij}{L}\sin\theta+\frac{j}{2L^2}[e^{i2\theta}-1\right.\nonumber\\
&&-(j-2)\sin^2\theta]+\frac{j}{2L^3}\left[e^{i\theta}-e^{i3\theta}-\frac{j-2}{2}(e^{i\theta}-e^{-i\theta})\right.\nonumber\\
&&\quad\left.\left.\times\left(e^{i2\theta}-1+\frac{j-4}{12}(e^{i\theta}-e^{-i\theta})^2\right)\!\right]\!\right\}\!
dL\, d\theta. \label{a2_5}
\end{eqnarray}
For $j=3$, the first integral is $-3\ln\pi/(2\pi n)$ and the second integral is $\sqrt{\pi}n^{-3/2}/8$. For $j=1$, the second integral yields $n^{-1}\int_\Lambda^\infty e^{-L^2/n} dL=(1/2)\sqrt{\pi/n}$. Thus, for odd $j$,
\begin{equation}
\mathcal{J}(n,j)=\frac{1}{2}\sqrt{\frac{\pi}{n}}\delta_{j1}-\frac{3\ln\pi}{2\pi n}\delta_{j3} + O(n^{-1}), \label{a2_6}
\end{equation}
as $n\to\infty$. For even $j$, the integrals over $\theta$ can be transformed into integrals over the unit circle on the complex plane and calculated by the residue theorem.
\begin{eqnarray*}
\int_{-\pi}^\pi \!\left(\frac{1+L e^{i\theta}}{1+L e^{-i\theta}}\right)^{j/2}\!\frac{d\theta}{\pi}=\int_{|z|=1}\! z^{\frac{j}{2}-1}\!\left(L+\frac{1-L^2}{z+L}\right)^{j/2}\!\frac{dz}{i\pi}.\quad 
\end{eqnarray*}
For $j=2$, the residue theorem yields
\begin{eqnarray*}
\int_{-\pi}^\pi \!\frac{1+L e^{i\theta}}{1+L e^{-i\theta}} \frac{d\theta}{\pi}=2(1-L^2)\,\Theta(1-L^2), 
\end{eqnarray*}
because the pole $z=-L$ is outside the unit circle if $|L|>1$. Here $\Theta(x)$ is the Heaviside unit step function. Then, as $n\to\infty$, we get
\begin{eqnarray}
\mathcal{J}(n,2)=\frac{2}{n}\int_0^1(1-L^2) L e^{-L^2/n}dL=\frac{1}{2n}+O(\frac{1}{n^{2}}). \label{a2_7}
\end{eqnarray}
Similarly, for $j=4$, the residue theorem yields
\begin{eqnarray*}
\int_{-\pi}^\pi \!\left(\frac{1+L e^{i\theta}}{1+L e^{-i\theta}}\right)^2\! \frac{d\theta}{\pi}=2(1-L^2)(1-3L^2)\,\Theta(1-L^2). 
\end{eqnarray*}
Then, as $n\to\infty$, we get
\begin{eqnarray}
\mathcal{J}(n,4)\!&=&\!\frac{2}{n}\int_0^1(1-L^2)(1-3L^2) L e^{-L^2/n}dL\nonumber\\
\!&=&\!\frac{1}{12n^2}+ O(\frac{1}{n^3}). \label{a2_8}
\end{eqnarray}
According to Eq.~\eqref{eq21},
\begin{widetext}
\begin{eqnarray}
C^{(1)}[e^{ij\theta}]_j = \!\left(\int^{\pi}_{-\pi}\! e^{-ij\xi} g(\xi)\,d\xi\right)\!e^{-M}\sum_{n=1}^\infty\frac{M^{n-1}}{(n-1)!}\!\!\left[(n-1)\frac{2J_1(|\mathbf{K}|R_0)}{|\mathbf{K}|R_0}+1\right]\mathcal{J}(n,j).\label{a2_9}
\end{eqnarray}
For large $M$, we expand $(n-1)\mathcal{J}(n,j)$ and $\mathcal{J}(n,j)$ about $M$ in this expression, thereby getting
\begin{eqnarray}
C^{(1)}[e^{ij\theta}]_j \sim \!\left(\int^{\pi}_{-\pi}\! e^{-ij\xi} g(\xi)\,d\xi\right)\!\!\left[(M-1)\frac{2J_1(|\mathbf{K}|R_0)}{|\mathbf{K}|R_0}+1\right]\mathcal{J}(M,j).\label{a2_10}
\end{eqnarray}
\end{widetext}
Then the functions $\mathcal{J}(n,j)$ produce Eqs.~\eqref{eq23} and 
\begin{eqnarray}\nonumber
Q_2=C^{(2)}[e^{i2\theta}]_2\sim\frac{1}{2}\int_{-\pi}^\pi e^{-i2\xi}g(\xi)d\xi,\\
Q_3=C^{(2)}[e^{i3\theta}]_3\sim-\frac{3\ln\pi}{2\pi}\int_{-\pi}^\pi e^{-i3\xi}g(\xi)d\xi,\nonumber\\
Q_4=C^{(2)}[e^{i4\theta}]_4\sim\frac{1}{12\, M}\int_{-\pi}^\pi e^{-i4\xi}g(\xi)d\xi. \label{a2_11}
\end{eqnarray}
Other eigenvalues tend to zero as $M\to\infty$. As $|\int_{-\pi}^\pi e^{-ij\xi}g(\xi)d\xi|\leq\int_{-\pi}^\pi g(\xi)d\xi=1$, the multipliers $Q_j$ with $j>1$  have moduli smaller than 1 in the limit as $M\gg 1$.
\bigskip

\section{Coefficients in the amplitude equations}\label{ap:1}
We calculate the coefficients in the amplitude equations by identifying them with others computed in \cite{ihl16}. Using Ihle's notation, we obtain $C^{(1)}[e^{i2\theta}]_2=Q_2=p$, $C^{(2)}[e^{i\theta},e^{i\theta}]_2=2\pi q$, $C^{(2)}[e^{-i\theta},e^{i2\theta}]_1=\pi S$, $C^{(3)}[e^{i\theta}\!,e^{i\theta}\!,e^{-i\theta}]_1=4\pi^2\Gamma$, \cite{ihl16}. In the limit as as $M\gg 1$, these identifications allow us to obtain the coefficients in Eqs.~\eqref{eq41} and \eqref{eq43}-\eqref{eq44}:
\begin{eqnarray}
Q_\eta\sim - \frac{\sqrt{\pi M}}{2\eta_c}\!\left( \frac{2}{\sqrt{\pi M}}- \cos\frac{\eta_c}{2}\right)\!, \label{a1_1}\\
\delta\sim\frac{2\gamma_0-1}{8}, \quad\gamma_0=\frac{1}{1-\frac{1}{\sqrt{\pi M}}\cos\frac{\eta_c}{2}},\label{a1_2}\\
\mu\sim\frac{\pi^4R_0^4/M}{1-\frac{1}{\sqrt{\pi M}}\cos\frac{\eta_c}{2}},\quad
\gamma_2\sim\frac{\gamma_0\pi^2R_0^2}{4M}. \label{a1_3}
\end{eqnarray}
To calculate the other coefficients appearing in the amplitude equations, we note that, as $M\gg 1$,
\begin{eqnarray}
C^{(2)}[1,e^{i\theta}]_1\!&=&\!\frac{\pi M}{\rho_0}\frac{\partial C^{(1)}[e^{i\theta}]_1}{\partial M}\nonumber\\
&\sim&\!\frac{\pi C^{(1)}[e^{i\theta}]_1}{2\rho_0}=\frac{\pi}{2\rho_0},\label{a1_4}\end{eqnarray}
\begin{eqnarray}
C^{(3)}[1,1,e^{i\theta}]_1\!\!&=&\!\frac{2\pi^4\! R_0^4}{3}\frac{\partial^2C^{(1)}[e^{i\theta}]_1}{\partial M^2}\nonumber\\
&\sim&\!-\frac{\pi^2C^{(1)}[e^{i\theta}]_1}{6\rho_0^2}=-\frac{\pi^2}{6\rho_0^2}. \label{a1_5}
\end{eqnarray}
We have calculated these coefficients at the critical noise $\eta_c$, where $C^{(1)}[e^{i\theta}]_1=1$. Eqs.~\eqref{a1_4} and \eqref{a1_5} yield the remaining coefficients:
\begin{eqnarray}
&&\Gamma(r)\sim\eta_2Q_\eta-\frac{r^2}{6\rho_0^2},\quad\gamma_3\sim\frac{\pi^2R_0^2}{2M}  \label{a1_6}\\
&&\gamma_1\sim\pi^2R_0^2\!\left(1+\frac{1}{8M}-\gamma_{0}\right)\!. \label{a1_7}
\end{eqnarray}

\section{Fourier coefficients $R_{n,m}(0)$}\label{ap:3}
Here we give examples of initial conditions used to calculate the solutions of Eqs.~\eqref{eq57}-\eqref{eq58}. A simple initial condition is to set $\mathbf{\tilde{w}}(\mathbf{X},0)=\mathbf{0}$. Then Eqs.~\eqref{eq66} and \eqref{eq67} yield 
\begin{eqnarray}
\dot{R}_{n,m}(0)=0, \quad\mathbf{C}_{n,m}=\mathbf{0}, \label{a3_1}
\end{eqnarray}
whereas Eq.~\eqref{eq64} gives
\begin{widetext}
\begin{eqnarray}
\epsilon\mathbf{\tilde{w}}_{0,0}(T)\!=\!(e^{\frac{\epsilon\gamma_3 w_{0x}L}{\pi}}-1)(e^{\frac{\epsilon\gamma_3 w_{0y}L}{\pi}} -1)\sum_{n=-\infty}^\infty\sum_{m=-\infty}^\infty\frac{ \!\left(i\frac{2\pi}{L}(n,m)-\frac{\epsilon\gamma_3\mathbf{w}_0}{\pi}\right)\!R_{n,m}(0)\!\sin(\omega_{n,m}T)}{\omega_{n,m}\!\left(\frac{\epsilon\gamma_3w_{0x}L}{\pi}+i2\pi n\right)\!\!\left(\frac{\epsilon\gamma_3 w_{0y}L}{\pi}+i2\pi m\right)\! }.  \label{a3_2}
\end{eqnarray}
\end{widetext}
We now have to calculate $R_{n,m}(0)$. One possibility is to have a function $\tilde{r}(\mathbf{X},0)$ with finitely many harmonics. For example, harmonics $(\pm 1,0)$ and $(0,\pm 1)$. We obtain 
\begin{widetext}
\begin{eqnarray}
R_{n,m}(0)\!&=&\!2(e^{\frac{\epsilon\gamma_3 w_{0x}L}{\pi}}-1)(e^{\frac{\epsilon\gamma_3 w_{0y}L}{\pi}}-1)\!\left\{ \frac{(\epsilon\gamma_3 w_{0x}L-i2\pi^2n)\text{Re}\tilde{r}_{1,0}(0)+2\pi^2\text{Im}\tilde{r}_{1,0}(0)}{\!\left[\!\left(\frac{\epsilon\gamma_3w_{0x}\!L}{\pi}-i2\pi n\right)^2+4\pi^2\right](\epsilon w_{0y}L-i2\pi^2m)} \right.\nonumber\\
\!&+&\! \left.\frac{(\epsilon\gamma_3 w_{0y}L-i2\pi^2m)\text{Re}\tilde{r}_{0,1}(0)+2\pi^2\text{Im}\tilde{r}_{0,1}(0)}{\!\left[\left(\frac{\epsilon\gamma_3w_{0y}\!L}{\pi}-i2\pi m\right)^2+4\pi^2\right](\epsilon\gamma_3 w_{0x}L-i2\pi^2 n)}\right\}\!. \label{a3_3}
\end{eqnarray}
\end{widetext}

Another simple possibility is the initial condition $\tilde{r}_{n,m}(0)=(-1)^{n+m}(1-\delta_{n0}\delta_{m0})$. Then Eq.~\eqref{eq65} yields
\begin{eqnarray}
R_{n,m}(0)\!&=&\! (-1)^{n+m} e^{\frac{\epsilon\gamma_3 w_{0x}L}{2\pi}}e^{\frac{\epsilon\gamma_3 w_{0y}L}{2\pi}}\nonumber\\
\!&+&\!\frac{e^{\frac{\epsilon\gamma_3 w_{0x}L}{\pi}}-1}{2n\pi+i\frac{\epsilon\gamma_3 w_{0x}L}{\pi}}\,\frac{e^{\frac{\epsilon\gamma_3 w_{0y}L}{\pi}}-1}{2m\pi+i\frac{\epsilon\gamma_3 w_{0y}L}{\pi}}. \label{a3_4}
\end{eqnarray}

 \section{Calculation of the shift in the critical noise}\label{ap:4}   
According to Eq.~(42) of Ref.~\cite{ihl16}, we have to approximate better the sum 
\begin{eqnarray}
S(M)=e^{-M}\sum_{n=0}^\infty\frac{M^n}{n!} h(n), \label{a4_1} 
\end{eqnarray}
where $h(n)$ has a maximum at $n=M$. Expanding this function about its maximum and keeping four terms in the expansion, we obtain 
\begin{eqnarray}
S(M)\sim h(M)+\frac{M}{2}h''(M)+\frac{M}{6}h'''(M)  \nonumber\\
+ \frac{(1+3M)M}{24}h^{(4)}(M), \label{a4_2} 
\end{eqnarray}
in which we have summed the corresponding series. To write Eq.~\eqref{eq23}, we took into account only the first term of Eq.~\eqref{a4_2} with $h(n)=\sqrt{n}$. For this function, we get
\begin{eqnarray}
S(M)\sim \sqrt{M}\left(1-\frac{1}{8M}+\frac{1}{16M^2}-\frac{5(1+3M)}{128M^3}\right)\!, \label{a4_3} 
\end{eqnarray}
which produces Eq.~\eqref{eq70}. For $M=7$, keeping more term in the expansion of $h(n)$ does not change appreciably the numerical value of $\eta_c$. Other corrections could come from calculating the term of order $n^{-3/2}$ in Eq.~\eqref{a2_6} because $h(n)$ is proportional to $n\,\mathcal{J}(n,1)$.

We now calculate $\langle r^2(\mathbf{X},T)\rangle$. We can use the Parseval equality and Eq.~\eqref{eq63} to get 
\begin{eqnarray}
&&\langle\tilde{r}^2(\mathbf{X},T)\rangle\!=\!\lim_{\mathcal{T}\to\infty}\frac{1}{\mathcal{T}}\int_0^\mathcal{T}\sum_{n,m}|\tilde{r}_{n,m}(T)|^2 dT\nonumber\\
\!&&\quad=\!\frac{1}{2}\sum_{n,m}\!\left[|R_{n,m}(0)|^2+\frac{|\dot{R}_{n,m}(0)|^2}{\omega_{n,m}^2}\right]\!g(\mathbf{w}_0), \label{a4_4}\\
&&g(\mathbf{w}_0)=\delta_{\mathbf{w}_0,\mathbf{0}}+\delta_{w_{0x},0}(1-\delta_{w_{0y},0})\frac{1+e^{\frac{\gamma_3}{\pi}w_{0y}\epsilon L}}{2\frac{\gamma_3}{\pi}w_{0y}\epsilon L}\quad\nonumber\\
&&+ \delta_{w_{0y},0}(1-\delta_{w_{0x},0})\frac{1+e^{\frac{\gamma_3}{\pi}w_{0x}\epsilon L}}{2\frac{\gamma_3}{\pi}w_{0x}\epsilon L}+\frac{1}{4}(1-\delta_{w_{0x},0})\nonumber\\
&&\times (1-\delta_{w_{0y},0})\frac{(1+e^{\frac{\gamma_3}{\pi}w_{0y}\epsilon L})(1+e^{\frac{\gamma_3}{\pi}w_{0x}\epsilon L})}{\left(\frac{\gamma_3}{\pi}\epsilon L\right)^2w_{0x}w_{0y}}.\label{a4_5}
\end{eqnarray}
To get these expressions, we have used that: (i) the averages of $\cos^2(\omega_{n,m}T)$ and of $\sin^2(\omega_{n,m}T)$ are both $1/2$, that the average of $\cos(\omega_{n,m}T)\cos(\omega_{l,j}T)$ is zero unless $n=l$, $m=j$, etc.; (ii) the integrals
\begin{eqnarray}
&&\frac{1}{\epsilon^2L^2}\int_0^{\epsilon L}\int_0^{\epsilon L} e^{(\frac{\gamma_2}{\pi}\mathbf{w}_0-i\mathbf{k}_{n-l,m-j})\cdot\mathbf{X}}d\mathbf{X}\nonumber\\
&&=\frac{(e^{\frac{\gamma_2}{\pi}w_{0x}\epsilon L}-1)\, (e^{\frac{\gamma_2}{\pi}w_{0y}\epsilon L}-1)}{[\frac{\gamma_3}{\pi}w_{0x}\epsilon L-i2\pi (n-l)][\frac{\gamma_3}{\pi}w_{0y}\epsilon L-i2\pi (m-j)]},\quad\quad\label{a4_6}
\end{eqnarray}
(iii) the sums
\begin{eqnarray}
\sum_{n=-\infty}^\infty\frac{1}{\frac{\gamma_3^2}{\pi^2}w^2_{0x}\epsilon^2L^2+4\pi^2 (n-l)^2}\!=\!\frac{\pi\mbox{coth}\frac{\gamma_3w_{0x}\epsilon L}{2\pi}}{2\gamma_3w_{0x}\epsilon L}.\label{a4_7}
\end{eqnarray}

In Eq.~\eqref{a4_4}, we can use again the Parseval equality and then the Schwarz inequality to obtain
\begin{eqnarray}
\sum_{n,m}\! |R_{n,m}(0)|^2\!=\!\frac{1}{\epsilon^2L^2}\!\int_0^{\epsilon L}\!\!\int_0^{\epsilon L}\tilde{r}^2(\mathbf{X},0)e^{-2\!\frac{\gamma_3}{\pi}\mathbf{w}_0\cdot\mathbf{X}}d\mathbf{X}\quad\nonumber\\
\leq\sqrt{\langle\tilde{r}(\mathbf{X},0)^4\rangle \frac{1}{\epsilon^2L^2}\!\int_0^{\epsilon L}\!\!\int_0^{\epsilon L}e^{-4\!\frac{\gamma_3}{\pi}\mathbf{w}_0\cdot\mathbf{X}}d\mathbf{X}}.  \quad\label{a4_8}
\end{eqnarray}
We can calculate $r^4=\epsilon^4\langle\tilde{r}^4\rangle$ by means of the grand canonical ensemble as \cite{hua87}:
\begin{eqnarray}
\langle (N-\langle N\rangle)^4\rangle= \!\left(z\frac{\partial}{\partial z}\right)^4\!\ln\mathcal{Q}(v,\beta)+3\langle(N-\langle N\rangle)^2\rangle^2 \nonumber\\
= \frac{1}{\beta^4}\frac{\partial^4}{\partial\mu^4}\ln\mathcal{Q}(v,\beta)+3\langle(N-\langle N\rangle)^2\rangle^2,\quad \label{a4_9}
\end{eqnarray}
where $z=e^{\beta\mu}$ is the fugacity, and $v=1/\rho_0=V/N$ ($V=L^2$), $\beta$ and $\mu$ are the specific area, the reciprocal of temperature in enegy units and the chemical potential, respectively. In the grand canonical ensemble, $\ln\mathcal{Q}(v,\beta)=\beta VP$, where $P(v)$ is the pressure,
\begin{eqnarray}
\mathcal{Q}(v,\beta)=\sum_{N=0}^\infty z^N Q_N(v,\beta), \label{a4_10}
\end{eqnarray}
and $Q_N(v,\beta)$ is the partition function of the canonical ensemble. The average number of particles in a volume $V$ is the density of particles and $\langle(N-\langle N\rangle)^2\rangle$ is the fluctuation of the density. In terms of the grand partition function, they are
\begin{eqnarray}
\langle N\rangle=\frac{\sum_{N=0}^\infty N z^N Q_N(v,\beta)}{\sum_{N=0}^\infty z^{N} Q_N(v,\beta)}=\!\left(z\frac{\partial}{\partial z}\right)\!\ln\mathcal{Q}(v,\beta), \label{a4_11}\\
\langle(N-\langle N\rangle)^2\rangle= \!\left(z\frac{\partial}{\partial z}\right)^2\!\ln\mathcal{Q}(v,\beta). \label{a4_12}
\end{eqnarray}
Following Ref.~\cite{hua87}, we now define pressure and chemical potential in terms of the free energy per particle $a(v)$ (we ignore the temperature dependence):
\begin{eqnarray}
A(N,V,T)=N\, a(v), \quad P=-\frac{\partial a(v)}{\partial v}, \nonumber\\
\mu= a(v)+ Pv.  \label{a4_13}
\end{eqnarray}
Using Eq.~\eqref{a4_13}, Eq.~\eqref{a4_11} becomes
\begin{eqnarray}
&&\langle N\rangle=\!\left(z\frac{\partial}{\partial z}\right)\!\ln\mathcal{Q}(v,\beta)=\frac{1}{\beta}\, \frac{\partial}{\partial\mu}\ln\mathcal{Q}(v,\beta)\nonumber\\
&&=V\, \frac{\partial P}{\partial\mu}= V\, \frac{\frac{\partial P}{\partial v}}{\frac{\partial\mu}{\partial v}}=V\frac{\frac{\partial P}{\partial v}}{\frac{\partial a(v)}{\partial v}+P+v\frac{\partial P}{\partial v}}=\frac{V}{v},\quad\quad \label{a4_14}
\end{eqnarray}
which is indeed the average number of particles. Then $1/v=\langle N\rangle/V=\rho_0$. Similarly, Eq.~\eqref{a4_12} yields
\begin{eqnarray}
\langle(N-\langle N\rangle)^2\rangle= \frac{V}{\beta}\,\frac{\partial}{\partial\mu}\frac{1}{v}=-\frac{V}{v^2\beta}\frac{1}{\frac{\partial\mu}{\partial v}}=\frac{\langle N\rangle}{-\beta v^2\frac{\partial P}{\partial v}}. \quad\label{a4_15}
\end{eqnarray}
This is Eq.~(7.43) of Ref.~\cite{hua87}. The particles can be thought of as belonging to an ideal gas at the initial time, therefore $P=\rho_0/\beta$ and $-v^2\partial P/\partial v=\partial P/\partial\rho_0=1/\beta$. Eq.~\eqref{a4_15} becomes $\langle(N-\langle N\rangle)^2\rangle=\langle N\rangle$, and therefore
\begin{eqnarray}
\langle r^2\rangle=\frac{\rho_0}{V}= \frac{\rho_0^2}{N}\Longrightarrow\sqrt{\langle r^2\rangle}=\frac{\rho_0}{\sqrt{N}}, \label{a4_16}
\end{eqnarray}
where now $N$ is the total number of particles in the box.

Using Eq.~\eqref{a4_15} and the ideal gas assumption in Eq.~\eqref{a4_15}, Eq.~\eqref{a4_9} becomes 
\begin{eqnarray}
&&\langle (N-\langle N\rangle)^4\rangle=\langle N\rangle+3\langle N\rangle^2 \Longrightarrow\nonumber\\ 
&&\langle r^4\rangle= \frac{3\rho_0^4}{N^2}\!\left(1+\frac{1}{3N}\right)\!.\label{a4_17}
\end{eqnarray}
Then, near the bifurcation point where $\mathbf{w}_0\epsilon L\ll 1$, Eq.~\eqref{a4_8} produces
\begin{eqnarray}
\epsilon^2\sum_{n,m} |R_{n,m}(0)|^2 \leq\frac{\rho_0^2\sqrt{3}}{N}\sqrt{1+ \frac{1}{3N}}.  \label{a4_18}
\end{eqnarray}
Let us assume that Eq.~\eqref{a4_18} is an equality and that $\dot{R}_{n,m}(0)=0$, which is the case if the initial current density is zero. Then Eq.~\eqref{a4_4} gives
\begin{eqnarray}
\langle r^2(\mathbf{X},T)\rangle\approx\frac{\sqrt{3}}{2N}\rho^2_0\sqrt{1+\frac{1}{3N}}, \label{a4_19}
\end{eqnarray}
and the shift in Eq.~\eqref{eq69} is 
\begin{eqnarray}
a^2=\frac{\langle r^2(\mathbf{X},T)\rangle}{6\rho_0^2}\approx\frac{\sqrt{3}}{12N}\sqrt{1+\frac{1}{3N}}. \label{a4_20}
\end{eqnarray}
This produces $a=0.01$ which is about a factor $30$ smaller than the shift in the critical noise measured from direct numerical simulations of the VM.

\end{document}